\documentclass[11pt]{article}

\usepackage{amsmath, amsthm, amsfonts, commath}
\usepackage{newtxtext,newtxmath}
\usepackage{bbm}
\usepackage{bm}
\usepackage{booktabs}
\usepackage{color}
\usepackage{dcolumn} 
\usepackage{epigraph}
\usepackage{enumitem}
\usepackage{epstopdf}
\usepackage{graphicx}
\usepackage{longtable} 
\usepackage{multirow}
\usepackage{natbib}
\usepackage{rotating}
\usepackage{setspace} 
\usepackage{subcaption}
\usepackage{placeins}
\usepackage[margin=1in]{geometry}
\usepackage{cancel}
\usepackage{physics}
\usepackage{tikz}
\usetikzlibrary{positioning, shapes.geometric, arrows.meta}
\PassOptionsToPackage{hyphens}{url}\usepackage{hyperref}
\hypersetup{
  colorlinks=true,
  citecolor  =[HTML]{8B0000},
  linkcolor  =[HTML]{8B0000},
  urlcolor   =[HTML]{8B0000}
}

\newcommand{\starlanguage}{Significance Indicator: $^{*}$p$<$0.05.}

\newcommand{\aisub}{AI agent}

\newcommand{\WorkingTitle}{Large Language Models as Simulated Economic Agents:\\ What Can We Learn from \emph{Homo Silicus}?}

\begin{document}  
\date{\today}
\title{\WorkingTitle{}\thanks{
Thanks to the MIT Center for Collective Intelligence, Tyler Cowen, and the Mercatus Center for their generous funding.
Thanks to 
Daron Acemoglu,
David Autor,
Mohammed Alsobay,
Jimbo Brand,  
Elliot Lipnowski,
Shakked Noy, 
Paul Röttger,
Daniel Rock, 
Charlotte Siegmann, and
Hong-Yi TuYe
for their helpful conversations and comments.
Special thanks to Yo Shavit, who has been extremely generous with his time and thinking.
All authors contributed equally to this work.
Author contact information, code, and data are currently or will be available at \url{https://benjaminmanning.io}.
Horton is an economic advisor to Anthropic;
all authors have a financial interest in \url{www.expectedparrot.com}.
In preparing this paper, the authors utilized generative AI models extensively as tools to assist with editing and evaluation. 
The authors retain full responsibility for all content and conclusions presented herein.
}
} 
\author{
John J. Horton      \\ MIT \& NBER  \and 
Apostolos Filippas  \\ Fordham      \and
Benjamin S. Manning \\ MIT
}

\maketitle
\doublespacing
\begin{abstract}
\noindent 
We argue that newly-developed large language models (LLMs), because of how they are trained and designed, are implicit computational models of humans---a \emph{Homo silicus}.
LLMs can be used like economists use \emph{Homo economicus}: they can be given endowments, information, preferences, and so on, and then their behavior can be explored in scenarios via simulation.
Experiments using this approach, derived from \cite{charness2002understanding}, \cite{kahneman1986fairness}, \cite{samuelson1988status}, \cite{Oprea2024lottery}, and \cite{laborHorton2025}, show qualitatively similar results to the original, and when they differ, it is often generative for future research.
We discuss potential applications, conceptual issues, and why this approach can inform the study of humans.
\end{abstract} 

\newpage \clearpage

\section{Introduction}
Most economic research takes one of two forms: (a) ``What would \emph{Homo economicus} do?'' and (b) ``What did \emph{Homo sapiens} actually do?'' 
The (a)-type research takes a maintained model of humans, \emph{Homo economicus}, endows it with different resources, preferences, information, and so on, subjects it to various economic scenarios, and then deduces its behavior.
This behavior can then be compared to that of actual humans in (b)-type research.

In this paper, we argue that large language models (LLMs)---because of how they are trained and designed---can be thought of as implicit computational models of humans---a \emph{Homo silicus}.
These models can be used the same way economists use \emph{Homo economicus}: they can be given endowments, put in scenarios, and then their behavior can be explored---though in the case of \emph{Homo silicus}, through computational simulation, not a mathematical
deduction.\footnote{\cite{lucas1980} writes, 
``One of the functions of theoretical economics is to provide fully articulated, artificial economic systems that can serve as laboratories in which policies that would be prohibitively expensive to experiment with in actual economies can be tested out at much lower cost.''}
This is possible because LLMs can now respond realistically to a wide range of textual inputs, giving responses similar to what we might expect from a human \citep{aher2022using,bowman2023things}.

Like all models, any particular \emph{Homo silicus} is wrong, but that judgment is separate from a decision about usefulness.
Of course, each \emph{Homo silicus} is a flawed model and can often give responses far away from what is rational or even sensical.
Yet unlike mathematical economic models with fixed input parameters, AI agents can be flexibly applied to simulate any scenario expressed in natural language.
Such flexibility means that agent fidelity to human responses is not fixed: the underlying instructions or ``prompts,'' which control their behavior can always be adjusted---even in ways that incorporate conventional economic theory.
Indeed, recently developed methods can dramatically improve agent fidelity, and as we will show, the process of better matching human responses can itself be generative.

We consider the reasons why AI simulations might be helpful in understanding actual humans.
The core of the argument is that LLMs---by nature of their training and design---are (i) computational models of humans and (ii) possess a great deal of social information.
For (i), the creators of LLMs have designed them to respond in ways similar to how a human would react to prompts \citep{bai2022helpfulharmless,ouyang2022training}---including prompts that are economic scenarios.
For (ii), these models capture two rich sources of social information.
The first is latent in the form of economic activities, decision-making heuristics, and common social preferences that the LLM implicitly learns from its training data.
These massive corpora contain a great deal of written text where people reason about and discuss economic matters: what to buy, how to bargain, how to shop, how to negotiate a job offer, how many hours to work, what to do when prices increase, and so on.
Second, LLMs have also been trained on the majority, if not all, of published social scientific research.
This comprises the facts, theory, and empirical results codified by researchers; welfare theorems, prospect theory's account of loss aversion, the income returns to education, and so on.

It is essential to note that the sources of social information in the LLM's training data are not perfect ingredients for simulating human behavior.
For instance, directly applying insights from published research may enhance the realism of simulations.
But it can also make brittle simulations appear robust because models may simply regurgitate problematic and memorized results \citep{sarkar2024lookahead,ludwig2025llmapplied}.
We address such tensions along with other conceptual issues related to LLMs' training corpora, opacity, and the generalizability of simulations.

While conceptual issues are important, what ultimately matters is whether these AI simulations are practically valuable for generating insights.
As such, the majority of this paper is devoted to presenting and discussing five simulated experiments.
We selected these experiments because they are simple to describe and implement.
They also have clear qualitative results that can be compared to the AI outcomes.

We begin with experiments motivated by \cite{kahneman1986fairness}, which reports survey responses to economic scenarios.
In the paper, there is an example where subjects imagine a hardware store raising the price of snow shovels following a snowstorm, and are asked to state whether doing this was fair or unfair.
Illustrating a benefit of \aisub{s}, unlike \cite{kahneman1986fairness}, we also vary the amount by which the store increases the price, and the political leanings of the respondent.
We show that large gouging is viewed more negatively and that endowed political views matter, with predictable effects---right-leaning AI agents are more sanguine about gouging.
The effects are robust to dozens of permutations of the original experiment.

Next, we simulate the simple unilateral dictator games from \cite{charness2002understanding}.
We show that endowing AI agents with various social preferences affects play.
Instructing the AI agent that it only cares about equity causes it to choose the equitable outcomes; telling the agent it cares about efficiency causes the selection of the payoff maximizing outcomes; telling the agent it is self-interested causes it to select allocations that maximize narrow self-interest.
Interestingly, without any endowment, some models choose efficiently while others tend to be selfish.
We use these results to generate new samples of agents that respond increasingly like human subjects in more complicated two-stage games.

In a study of framing, we next present the AI agents with a decision-making scenario introduced by \cite{samuelson1988status}.
In the paper, the respondent has to allocate a federal budget between highway safety and car safety.
The original results show that humans are subject to a status quo bias, preferring budget options that are presented as the status quo.
We recapitulate this result by endowing AI agents with baseline views about the relative importance of car or highway safety. 
We then put those agents through various scenarios, with each possible allocation taking a turn as the status quo.
Some LLMs exhibit a strong bias towards any option presented as the status quo, while others barely do.

In the fourth set of simulations, we explore the experiments in \cite{Oprea2024lottery}, where participants were asked to supply certainty equivalents for a series of binary choice lottery-type questions.
This research challenges conventional interpretations of prospect theory by showing that the classic ``fourfold pattern'' of risk attitudes emerges even when participants face no risk. 
Rather, the paper posits that the pattern emerges because cognitively constrained participants are heuristically responding to complexity.
We simulate the experiment across an expanded set of questions with five different models serving as the AI agents.
Even though the paper is recent and contrasts decades of research, our simulated results provide suggestive evidence in support of \citeauthor{Oprea2024lottery}'s complexity-driven interpretation.

Finally, we explore a hiring scenario motivated by \cite{laborHorton2025}, which shows in a field experiment that employers facing a minimum wage substitute towards higher-wage workers. 
We create a scenario where an employer is trying to hire a worker as a dishwasher and faces a collection of applicants that differ in their experience and wage asks, which are chosen randomly.
The AI agent makes an experience-wage tradeoff.
We then impose a minimum wage that forces applicants bidding below that minimum to bid up.
We run these scenarios across LLMs with varying capabilities.
On average, the results are directionally consistent with \citeauthor{laborHorton2025}: the minimum wage increase causes a shift in the AI's hiring of more experienced applicants and higher hourly wages.

The first draft of this paper featured versions of the now-deprecated \textsc{Gpt-3} as the underlying models in our experiments.
Results from these now woefully out-of-date models appear in Appendix~\ref{sec:original_figures}.
We now employ dozens of contemporary LLMs---both open-source and proprietary---and introduce methodological improvements that can increase realism and test robustness of the simulations.
However, as newer models are inevitably released, even our updated experiments will become dated. 
Because this pattern will continue to repeat, and access to proprietary systems is uncertain, we emphasize the importance of simulations that are easy to design, replicate, and scale.

This paper is now accompanied by an open-source Python package that enables researchers to replicate our experiments on new models or explore their own ideas with minimal technical overhead, often using only a fraction of the code required for the simulations in the initial draft.\footnote{See \url{https://pypi.org/project/edsl/} to download and \url{https://docs.expectedparrot.com/} for documentation.}
Using this package, all experiments in this paper are now button-push reproducible with almost any LLM, and the software is actively maintained as methods, models, and prompts improve.
We include links to code, prompts, and tutorials for all simulations throughout the manuscript.
Whether or not researchers choose to employ this software, or something else, it is essential---and now feasible---for AI simulations to have the features we have highlighted in the package: to be nearly costless to replicate, share, and update.
These are precisely the features social scientists should desire from any experimental method.

Ultimately, we care about the behavior of actual humans, and so results from AI experiments will still require empirical confirmation \citep{ludwig2025llmapplied}.\footnote{
  Our focus is not on using LLMs as experimental subjects or studying their behavior directly, though both are valuable research directions \citep{alberti2019bert, westby2022,binz2023turning}, but on using them as tools to understand humans.
}
As such, what is the value of these experiments? 
The most obvious use is to pilot experiments in silico first to gain insights.
They could cheaply and easily explore a parameter space, test whether behaviors seem sensitive to the precise wording of various questions, generate data that will ``look like'' the actual data, and inform power calculations.\footnote{
  There is an analogy to protein-folding, where simulations identify proteins later found in the real world \citep{kuhlman2003design}.
  }
The advantages in terms of speed and cost are enormous---the experiments in this paper were run in minutes for a trivial amount of money.
As relationships are identified, they could guide actual empirical work---or interesting effects could be captured in more traditional theory models.
Indeed, by endowing AI agents with prompts grounded in social science theory, their predictive capabilities can improve dramatically \citep{manning2025general}---precisely what researchers seek from a good theory.
This use of simulation as an engine of discovery is similar to what many economists do when building a ``toy model''---a tool not meant to be reality but rather a tool to help us think and identify promising patterns.

Since the first draft of this paper in 2023, an extensive and rapidly growing literature testing \aisub{s} has emerged.
This work, including thousands of experiments, now spans every social scientific discipline \citep{brand2023using, chang2024networks, zhu2025evidence,wang2025market, fish2025collusion, qian2025strategicbargaining}, with some studies building on these simulations to study humans in ways previously infeasible \citep{Manning2024Automated, Jackson2025Mixture, bybee2025ghost}. 
In a growing number of cases, AI agents put in simulations of the relevant settings can even predict human behavior in never-before-seen surveys and experiments \citep{ValidityLLM2024, hewitt2024predictingLLM, LLMtheory2024Tranchero, 1000agents2024park, binz2025centaur, suh2025finetuningscaled}.

Given this rapid expansion, the contributions of this paper are now twofold. 
First, we aim to help economists effectively incorporate AI simulations into their methodological toolkit, utilizing the most up-to-date practices, and ensuring these tools remain useful as models and methods continue to evolve. 
Second, the relative---and original---contribution of this paper is drawing the connection to the common research paradigm of economics and the role a foundational assumption (or model) like rationality plays in research. 
LLM experimentation is more akin to the practice of economic theory, despite superficially looking like empirical research.

The remainder of this paper is structured as follows.
In Section~\ref{sec:experiments}, we describe and analyze five simulated experiments.
Section~\ref{sec:when-why} discusses why LLMs can often predict human behavior and when they are likely to be useful.
In Section~\ref{sec:conceptual_issues}, we address conceptual critiques related to the use of LLMs for social science.
The paper concludes in Section~\ref{sec:conclusion}.

\section{Experiments}
\label{sec:experiments}
We conduct five experiments with LLMs.
We recapitulate and extend four laboratory experiments and then one field experiment from the economics literature.
Our use of the term ``recapitulate'' in lieu of ``replicate'' is intentional: we do not view LLM experiments as direct copies of the originals.
The goal is not to just reproduce results.
Rather, we aim to see how simulations behave in empirical scenarios where we already understand the analogous human behavior.
Then, by contrasting with the originals, we demonstrate key features of simulations that researchers can manipulate to better explore new ideas.

The first experiment demonstrates how we can use LLMs to easily extend experiments and search for new insights.
In the second, we show that LLMs respond reasonably to prompts and follow instructions effectively. 
We also show how to leverage these responses and conventional economic theory to construct samples of \aisub{s} that respond increasingly like human subjects.
The third experiment shows that LLMs exhibit well-established human decision-making biases. 
The fourth shows that LLMs can be used to study subtle elements of human decision-making and induce experimental variation in new ways.
The fifth experiment demonstrates that researchers can use \aisub{s} to study realistic field-like decision-making scenarios.

For readers unfamiliar with LLMs, we provide a brief primer in Appendix~\ref{app:primer}.
Other introductory resources include~\cite{bowman2023things} and~\cite{ouyang2022training}, while~\cite{zhao2025surveyllms} provide a comprehensive and technical survey of the field.
In addition, \cite{GenAIEcon2023Korinek} and \cite{Charness2025NextGen} examine how LLMs can be used across the full spectrum of experimental economics research.

\subsection{Fairness as a constraint on profit-seeking \citep{kahneman1986fairness}} 
\label{sec:kkt}
\cite{kahneman1986fairness} conducted experiments designed to study people's fairness perceptions in market transactions.
In their price gouging scenario, they asked subjects to respond to the following vignette:
\begin{quote}
A hardware store has been selling snow shovels for \$15. 
The morning after a large snowstorm, the store raises the price to \$20. 
Rate this action as: (1) Very Unfair, (2) Unfair, (3) Acceptable, (4) Completely Fair.
\end{quote}
About 82\% of subjects responded either ``Unfair''
or ``Very Unfair.''

We use \aisub{s} to explore two natural follow-up questions: 
(i) whether the subjects' political preferences and associated attitudes toward markets affect their views, and
(ii) whether there exists a dose-response relationship, with more aggressive price hikes seen as more egregious.
To do so, we create \aisub{s} endowed with ``personas'' or ``types,'' by prepending a description of the persona in the prompt.
We endow \aisub{s} with political views ranging from ``You are a socialist'' to ``You are a libertarian,'' and ask each to independently assess the fairness of price increases to \$16, \$20, \$40, and \$100.
We use \textsc{Gpt-4} with the model temperature at 1, and collect 100 responses to each combination of price hikes and political views, which generates 2,400 observations.

\begin{figure}[h!]
\begin{center}
\begin{minipage}{1 \linewidth}
\caption{Recapitulation and extension of \cite{kahneman1986fairness}'s price gouging experiment.} 
\vspace*{-0.1in}
\label{fig:kkt}
\includegraphics[width = \linewidth]{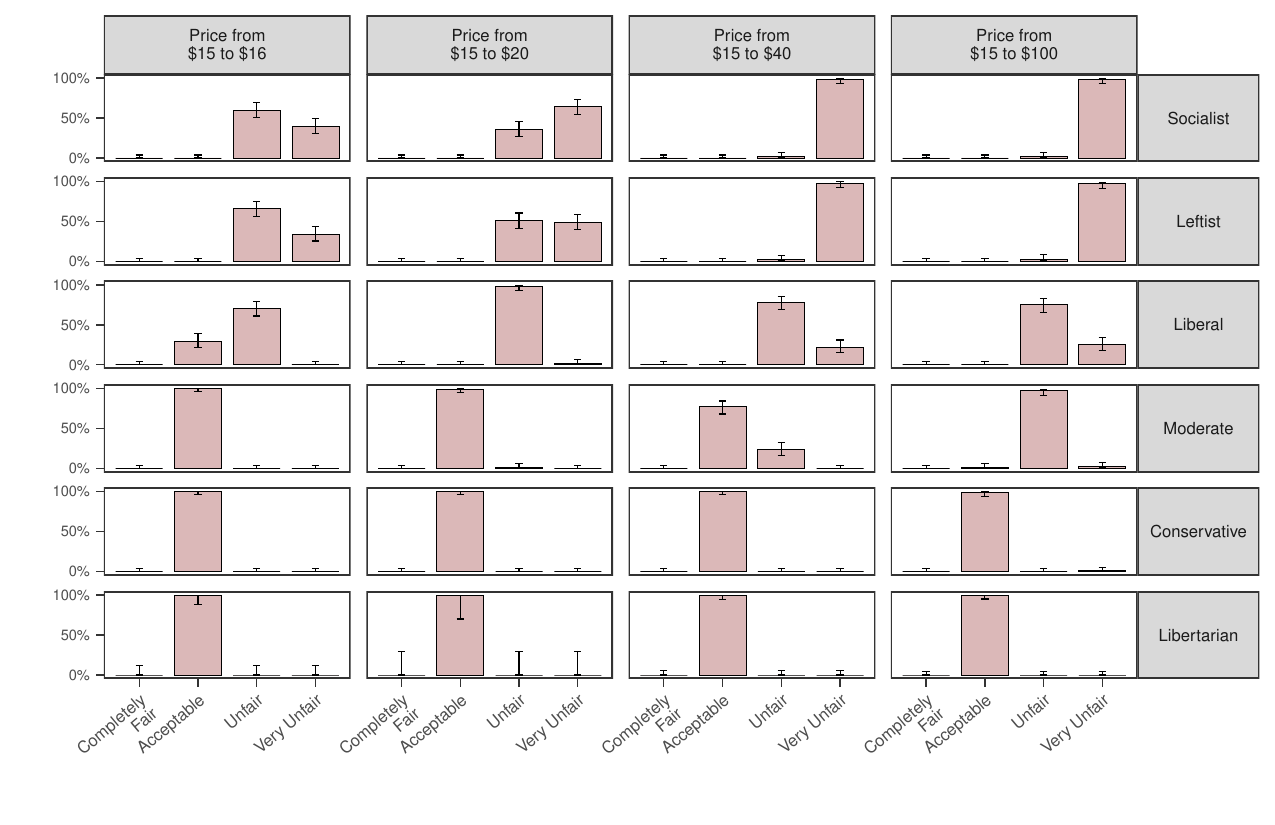}
\end{minipage}
\end{center}
\begin{footnotesize}
\begin{singlespace}
\vspace*{-0.35in}
\emph{Notes: }
This figure reports the results of our experiment recapitulating and extending the price gouging experiment from~\cite{kahneman1986fairness}. 
Each column corresponds to a different price hike scenario, and each row to the \aisub{s}' political leanings. 
The x-axis shows the possible responses, and the y-axis shows the proportion of \aisub{s} choosing each option.
Error bars represent 95\% multinomial Wilson confidence intervals.
\aisub{s} use \textsc{Gpt-4} with the temperature set to 1. 
We collect 100 responses for each \aisub{} persona and scenario combination. 
See \url{www.expectedparrot.com/content/0e39f555-fe34-4911-b516-46638d1640ca} for more details on prompt construction.
\end{singlespace}
\end{footnotesize}
\end{figure}

Figure~\ref{fig:kkt} reports the results of our experiment and details on how we constructed the prompts can be found in the notes via the linked Jupyter notebook.
To get a sense of the structure and content of the code, see Figure~\ref{fig:edsl_screenshot}, which provides a screenshot of the notebook in the appendix.
We provide such notebooks, which are accompanied by extensive documentation and instructions, for all experiments throughout this paper in their respective figure or table notes.
Columns of this figure correspond to different price increase scenarios and rows to different political bents of the \aisub{s}.
The x-axis shows the possible responses, and the y-axis shows the proportion of agents who choose each option.

The \aisub{s}' political beliefs affect their views about the price hikes: Socialist and Leftist agents judge the hikes to be ``Unfair'' or ``Very unfair,'' and subjects with more Conservative or Libertarian views find price hikes more morally permissible.
There is also a clear dose-response relationship, with larger price hikes generally viewed as less fair than smaller ones---even a few of the Conservative \aisub{s} judge the \$100 price hike as ``Unfair'' or ``Very Unfair''.
The answers of the \aisub{s} are also fairly consistent within political view-price gouge combinations: in about half of the panels, we get the same response in all 100 observations.

\citeauthor{kahneman1986fairness} report little about the composition of their original study, making it difficult to know which combination of political views is representative.
They do note that the sample consisted of Canadians from Toronto and Vancouver, with an approximately even split between males and females.
While we have no way to know the political leanings of the participants, it is plausible that the original sample was skewed left-leaning, given their urban origins.
Directionally consistent with this speculative view and the results in \citeauthor{kahneman1986fairness}, \aisub{s} endowed with left-leaning views always judge the original price hike to be ``Unfair'' or ``Very Unfair.''

\subsubsection{Testing generalizability and possible memorization}
\label{sec:robust}
One concern with our fairness simulations is that the \aisub{s} may have memorized both the original study and other texts describing the relationships between political labels and fiscal views.
If true, our results might be brittle---an artifact unique to the snow shovel example that does not generalize to other related scenarios.
We next show how to better evaluate the robustness of AI simulations.

\begin{figure}[h!]
\begin{center}
\begin{minipage}{1 \linewidth}
\caption{Coefficients of the effects of political beliefs on \aisub{s}' fairness assessments in four permutations of the \cite{kahneman1986fairness} experiment.} 
\vspace*{-0.1in}
\label{fig:kkt_coef_plot}
\includegraphics[width = \linewidth]{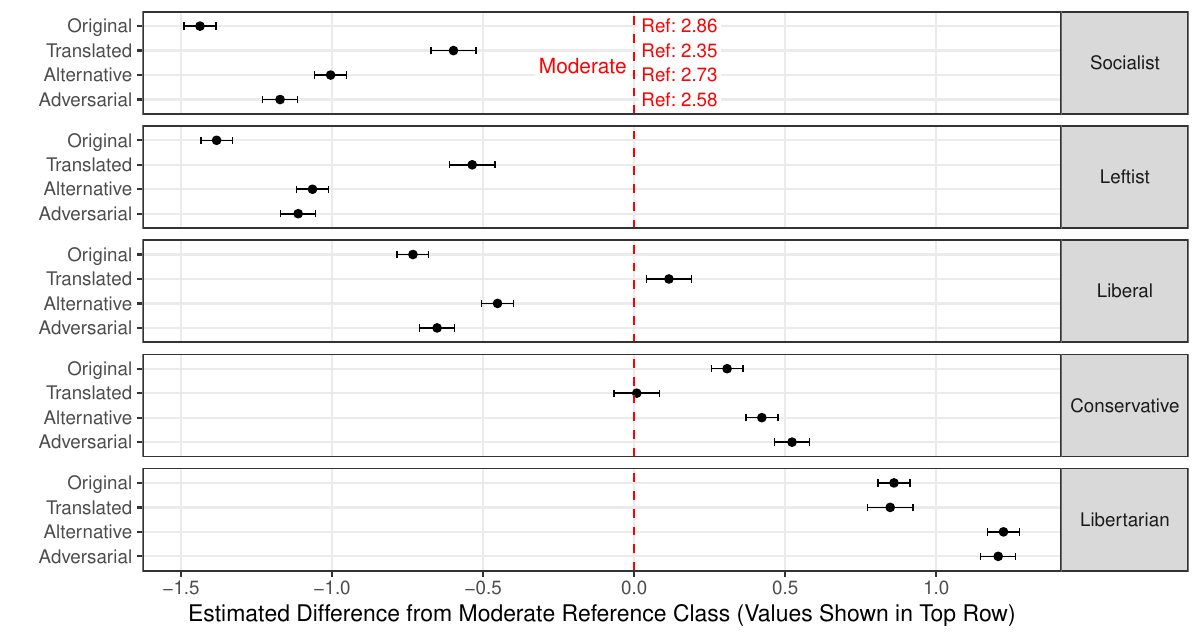}
\end{minipage}
\end{center}
\begin{footnotesize}
\begin{singlespace}
\vspace*{-0.15in}
\emph{Notes: }
This figure reports regression coefficients for political belief indicators, with Moderate as the omitted reference group, controlling for price gouging.
The dependent variable is the fairness rating (1 = Very Unfair, 4 = Completely Fair).
Full specifications appear in Table \ref{tab:kkt}.
The x-axis shows estimated differences from the Moderates; the y-axis shows the experimental permutation.
Each row corresponds to a different political belief.
The center of the top panel reports the reference intercepts for each permutation---the mean fairness rating for Moderates.
All other estimates are the difference from these values.
See \url{www.expectedparrot.com/content/6b9ca070-bf21-4887-b68d-060c9177650f} for simulation code and all experimental variations.
\end{singlespace}
\end{footnotesize}
\end{figure}

We conduct three follow-up sets of simulations of the price-gouging study.
First, we translate the original prompt into 10 different languages and run the experiment in each language.\footnote{These are French, German, Spanish, Italian, Portuguese, Greek, Japanese, Mandarin, Korean, and Arabic.}
Second, we ask \textsc{Gpt-4} to generate ten alternative versions of the original phrasing, varying the item being sold, the location, the event, or a combination of all three.
For the full list of alternatives, see the accompanying Jupyter notebook in the Figure~\ref{fig:kkt_coef_plot} notes.
One example is: 
\begin{quote}
  A bakery has been selling artisan bread loaves for \$15.
  The morning before a major holiday, the bakery raises the price to \$20.
  Rate this action as: (1) Very Unfair, (2) Unfair, (3) Acceptable, or (4) Completely Fair.
\end{quote}
Others include a bookstore selling novels after an author announces a signing event, a convenience store selling sunscreen right before a heatwave, etc.
Third, we ask GPT-4 to generate ten ``adversarial'' versions of the original phrasing, that is, to generate a vignette which would result in \aisub{s} responding differently than in our original experiment.
Specifically, we prompted the LLM to make the 10 adversarial versions ``as diverse as possible such that the same combination of variables will get people to answer differently.''
The adversarial examples are the same format as the alternative examples and are also provided in the same notebook.
We conduct all three of these additional experiments 10 times for each of the 10 prompt variations within every political view-price gouge combination, with the temperature parameter set to 1.
With four price gouges, six political leanings, and 10 variations for each permutation, we collect 2,400 observations for each of the additional experiments.

To measure responsiveness to price gouging, we regress the \aisub{s}' choices, measured as a continuous variable ranging from 1 (``Very Unfair'') to 4 (``Completely Fair''), on the price change and the \aisub{s}' political endowments.
Figure~\ref{fig:kkt_coef_plot} reports dummy regression coefficients for different political endowments across the four experimental permutations, using ``Moderate'' as the excluded reference category.
Each row represents a different political belief, with the x-axis showing estimated differences from the reference class.
Values for the reference OLS constant are shown in red in the top panel for each experimental permutation.

For interpretation, the translated permutation's Moderate intercept of 2.35 indicates that, on average, \aisub{s} endowed with moderate political views in this permutation rate price increases as roughly one third of the way from ``Unfair'' to ``Acceptable.'' 
The Socialist coefficient of -0.60 for the translated permutation in the top panel means Socialist \aisub{s}, on average, rate price increases as one quarter of the way from ``Unfair'' to ``Very Unfair'' (2.35 + (-0.60) = 1.75).

As in the original simulations, agents with more right-leaning political orientations generally rate price gouging as increasingly fair across all permutations.
This pattern is evident in the diagonal trend of coefficients from the upper left to the lower right. 
Notably, the adversarial and alternative examples yield strikingly similar results. 
We were unable to invert the relationship—agents with more left-leaning orientations never rated price gouging as fairer than those on the right. 
At a high level, the original pattern appears quite robust.

The translated permutation, however, exhibits a few small differences. 
Liberal \aisub{s} show positive coefficients, suggesting they are more right-leaning than the Moderate reference group. Conversely, the Conservative agents in this permutation appear nearly indistinguishable from the Moderates. This variation may reflect different definitions of ``Conservative'' and ``Liberal'' political views outside English-speaking contexts.\footnote{For example, in China: \url{en.wikipedia.org/wiki/Conservatism_in_China}.}

Although not visualized here, higher price increases are viewed as increasingly unfair, consistent with results in Figure~\ref{fig:kkt}. 
Table~\ref{tab:kkt} in the appendix reports the full specifications with these values.
Across the four permutations, including the original, a \$1 price hike causes between a 0.004 and 0.009 point decrease in the Likert fairness assessment.

\subsection{A social preferences experiment~\citep{charness2002understanding}} 
\label{sec:charness_rabin}
\cite{charness2002understanding} ask experimental subjects to choose between allocations that involve trade-offs between efficiency and equity.
We focus on their unilateral dictator games, where the dictator (Person B) chooses between two allocations of money between herself and the other player (Person A).
For example, in the following game:
\begin{quote}
\centering
\begin{tabular}{r l}
``Left'': & Person A gets 400 and Person B gets 600\\
``Right'': & Person A gets 700 and Person B gets 300,
\end{tabular}
\end{quote}
the dictator chooses between the ``Left'' and ``Right'' allocations.
We can write this game as:
\begin{quote}
\centering
  Person B Chooses: $\underbrace{(\underbrace{400}_{\mbox{to A}}\:, \:\underbrace{600}_{\mbox{to B}})}_{\mbox{``Left''}}$ $\quad$ vs. $\quad$ $\underbrace{(\underbrace{700}_{\mbox{to A}}\:, \:\underbrace{300}_{\mbox{to B}})}_{\mbox{``Right''}}$.
\end{quote}

We recapitulate this experiment using \aisub{s}.
First, we prompt LLMs with descriptions for each game, and ask them to respond with their preferred allocation without any additional information.
Second, we construct three \aisub{s} with the following personas: (i) inequity-averse (\emph{``You only care about fairness between players''}), (ii) efficient (\emph{``You only care about the total payoff of both players''}), and (iii) self-interested (\emph{``You only care about your own payoff''}).
We have each \aisub{} play each game 100 times with the temperature set to one.
We use \textsc{Gpt-4o}, \textsc{Claude-Sonnet-3.5}, \textsc{Llama-3-70B}, and \textsc{Deepseek} as the underlying models.

\begin{figure}[h!]
\begin{center}
\begin{minipage}{1 \linewidth}
\caption{Recapitulation of single-stage dictator games from~\cite{charness2002understanding}.} 
\vspace*{-0.1in}
\label{fig:charness_rabin}
\includegraphics[width = \linewidth]{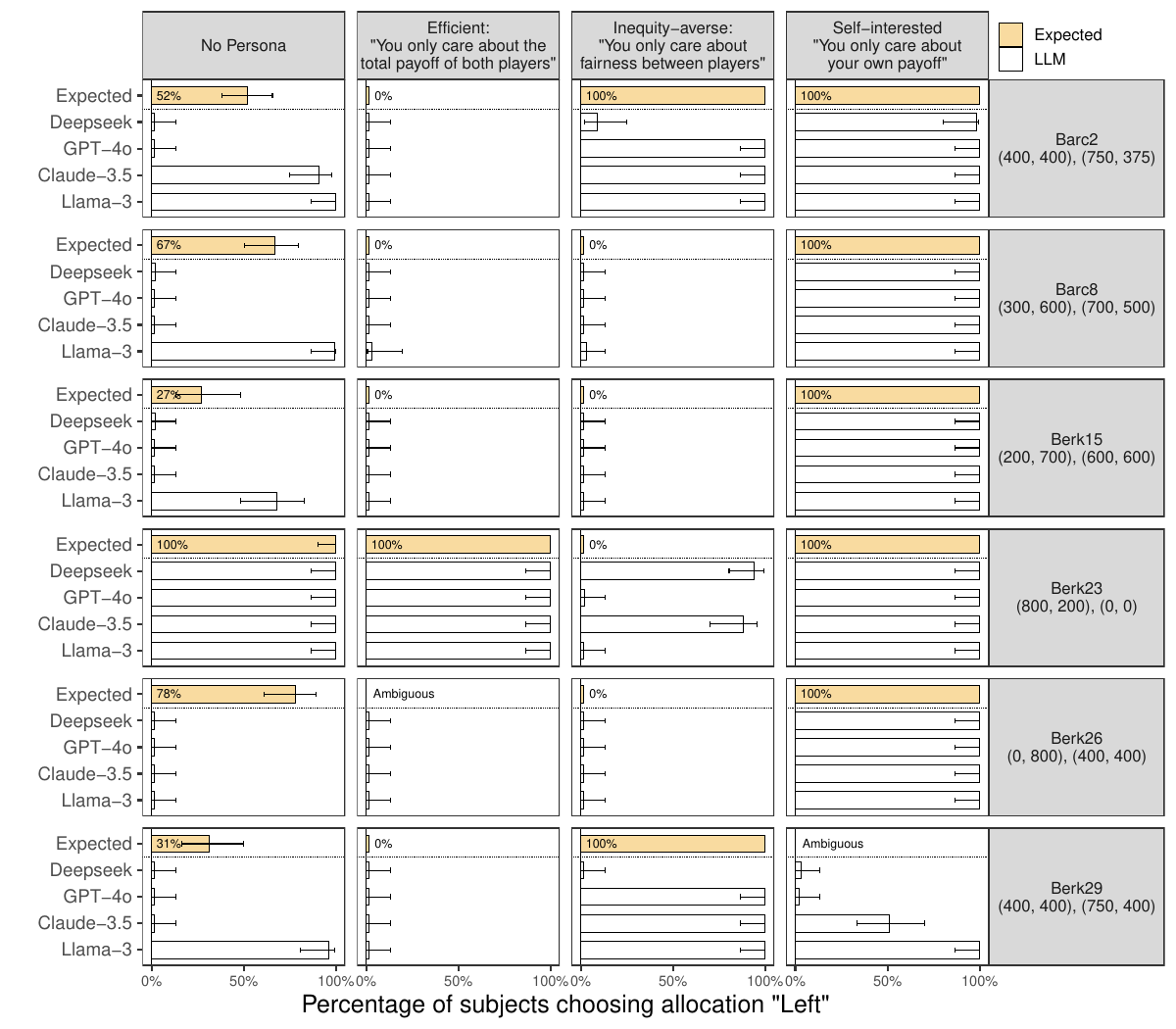}
\end{minipage}
\end{center}
\vspace*{-0.12in}
\begin{footnotesize}
\begin{singlespace}
\emph{Notes:}
This figure reports the recapitulation of unilateral dictator games from~\cite{charness2002understanding} with \aisub{s}.
The columns correspond to personas with which we endow \aisub{s}, and the rows to different games.
Each \aisub{} plays each game 100 times.
The x-axis shows the percentage of agents choosing ``Left.'' 
The y-axis shows both the \aisub{s}' responses and the ``Expected'' proportion for each column's given persona.
White bars depict the AI agents' choices, and gold bars depict either the choice of the human subjects in~\citeauthor{charness2002understanding} (leftmost column) or the choice that a perfectly aligned \aisub{} would make with the given persona (all other columns).
Error bars report 95\% Wilson confidence intervals.
See \url{www.expectedparrot.com/content/6bc88957-c346-46a3-910a-e63ec2d77035} for simulation walkthroughs.
\end{singlespace}
\end{footnotesize}
\end{figure}

Figure~\ref{fig:charness_rabin} plots the results of our experiment with the notebook to run the simulations linked in the notes.
Each column corresponds to a different persona, and each row to a different game.
The x-axis shows the fraction of agents choosing the ``Left'' option.
The white bars show the \aisub{s}' choices, and the gold bars show either the choice of the human subjects in~\citeauthor{charness2002understanding} (leftmost column) or the choice that a perfectly adherent \aisub{} would make with the given persona (all other columns).
For example, in the Barc2 game, 52\% of the human subjects chose ``Left'' (400,400), and the efficient choice, with the maximum total payout, would be to always select ``Right.''
  
The choices of the persona-less \aisub{s} are quite different from those of the human participants in the original experiment, despite the fact that the original experiment is almost certainly included in the training data for all models (as is true for most well-known economics papers).
These responses are also different across models, although consistent within: \textsc{Llama-3-70B} is seemingly more selfish, and \textsc{Gpt-4o}, \textsc{Claude-Sonnet-3.5}, and \textsc{Deepseek} are more efficiency-minded.
Despite this baseline heterogeneity, \aisub{s} endowed with personas respond consistently in line with those personas. 
Except for a few cases---such as \textsc{Deepseek} and \textsc{Llama-3-70B} failing to choose the inequity-averse option in the spiteful Berk23 scenario---efficiency-endowed agents generally chose the option with the highest combined payoff, inequity-averse agents selected the option with the smallest difference between payoffs, and self-interested agents maximized their own payoff.

\subsubsection{Constructing a calibrated sample of \aisub{s}} \label{sec:calibrate}

If we wanted to use these off-the-shelf LLMs without personas to predict behavior in new allocation games, the evidence above suggests their accuracy would be poor. 
Such agents might suffice for low-stakes applications---like stress-testing experimental designs or checking data pipelines---where fidelity to human distributions is not the primary concern. 
But if the goal is to employ AI agents as credible guides for empirical work---a far more valuable use case and promising application of AI simulations---then this lack of predictive reliability becomes a serious limitation.

By contrast, the \aisub{s} endowed with personas exhibited high fidelity: they followed the instructions tied to their assigned type almost perfectly.
This disciplined behavior provides a foundation for building calibrated samples of agents that can better predict human responses.
We adopt the approach from \cite{manning2025general}, who show that by optimizing mixtures of theory-grounded agents to match human responses in one setting, one can improve their predictive power in other related settings.
Here, ``theory-grounded'' means the personas are motivated by some theory that we generally understand, can clearly evaluate, and reasonably expect to predict the relevant human behavior---like how fairness, self-interest, and efficiency agents reflect the original economic model laid out in \citeauthor{charness2002understanding}.

We implement the approach and construct an optimized mixture of \aisub{s} as follows.
Intuitively, we choose the ``recipe'' of agent types---self-interested, inequity-averse, and efficient---whose combined choices best reproduce the aggregate choices of human subjects.
Each persona is represented by a vector $v = (p_{Barc2}, \ldots, p_{Berk29})$, where $p_g$ is the proportion of \aisub{s} with this persona that chose ``Left'' in game $g$ from Figure~\ref{fig:charness_rabin}.
For example, the ``efficient'' \textsc{Gpt-4o} \aisub{s} only chose ``Left'' in the Berk23 scenario, so they are represented by $v_{\text{E}}  = (0, 0, 0, 1, 0, 0)$; the population of~\citeauthor{charness2002understanding} is represented by $v_{CR} = (.52, .67, .27, 1, .78, .68)$.
For each LLM, we then compute the weights $w = (w_E,w_I, w_S)$ that minimize the mean-squared error (MSE) $||w_Ev_E + w_Iv_I + w_Sv_S - v_{CR}||^2$, subject to $w_E + w_I + w_S = 1$ and $w_E, w_I, w_S \geq 0$.\footnote{Others have explored this mixture method \citep{leng2024calibrate,Jackson2025Mixture,bui2025mixture}.}
The optimal weights for each LLM are $(0.49, 0.00, 0.51)$ for \textsc{Llama-3-70B}, $(0.44, 0.00, 0.56)$ for \textsc{Claude-Sonnet-3.5}, $(0.37, 0.10, 0.53)$ for \textsc{Gpt-4o}, and $(0.44, 0.00, 0.56)$ for \textsc{Deepseek}.
We use these weights to randomly sample personas for each LLM.

Having calibrated these mixture weights on the unilateral games, we now evaluate whether they generalize to new settings.
We use the two-stage games from~\citeauthor{charness2002understanding} for this out-of-sample evaluation. 
These games have a different structure, but are theoretically similar to the unilateral dictator games in the sense that we might reasonably expect some combination of efficiency, fairness, and self-interest to influence human responses in both.
If our calibrated agents can better predict responses in these new games than the persona-less LLMs, this provides evidence that they can generalize across structurally distinct data-generating processes.

In the sequential games, monetary allocations depend on both players' decisions.
In the first stage, Person A chooses either a given allocation or to let Person B choose one of two other known allocations.
Person B is asked to choose an allocation but is not informed of Person A's choice---until the payoffs are realized.
For example,
\begin{quote}
Stage 1 (Person A chooses): $\underbrace{(\underbrace{500}_{\mbox{To A}}\:, \:\underbrace{500}_{\mbox{To B}})}_{\mbox{``Left''}}$ $\quad$ vs. $\quad$  $\underbrace{\underbrace{(400\:\:,\:600)\:\text{    vs.  }(700\:\:,\: 300)}_{\mbox{Let Person B choose}}}_{\mbox{``Right''}}$.
\end{quote}
\begin{quote}
Stage 2  (Person B chooses): $\underbrace{(\underbrace{400}_{\mbox{To A}}\:,\: \underbrace{600}_{\mbox{To B}})}_{\mbox{``Left''}}$ $\quad$ vs. $\quad$ $\underbrace{(\underbrace{700}_{\mbox{To A}}\:,\: \underbrace{300}_{\mbox{To B}})}_{\mbox{``Right''}}$.
\end{quote}
We write this game as:
\begin{center}
  (500, 500) \\
  (400, 600), (700, 300)
\end{center}

We conduct our experiment as follows. 
For each LLM, we sample randomly 100 \aisub{s} according to the computed weights $w$.
Each \aisub{} plays each game scenario 5 times, both as Person A and Person B.
We also have ``control'' persona-less \aisub{s} play each game and role 100 times.

\begin{figure}[h!]
\begin{center}
\begin{minipage}{1 \linewidth}
\caption{Out-of-sample evaluations of calibrated samples of \aisub{s}.} 
\vspace*{-0.1in}
\label{fig:charness_rabin_mixture}
\includegraphics[width = \linewidth]{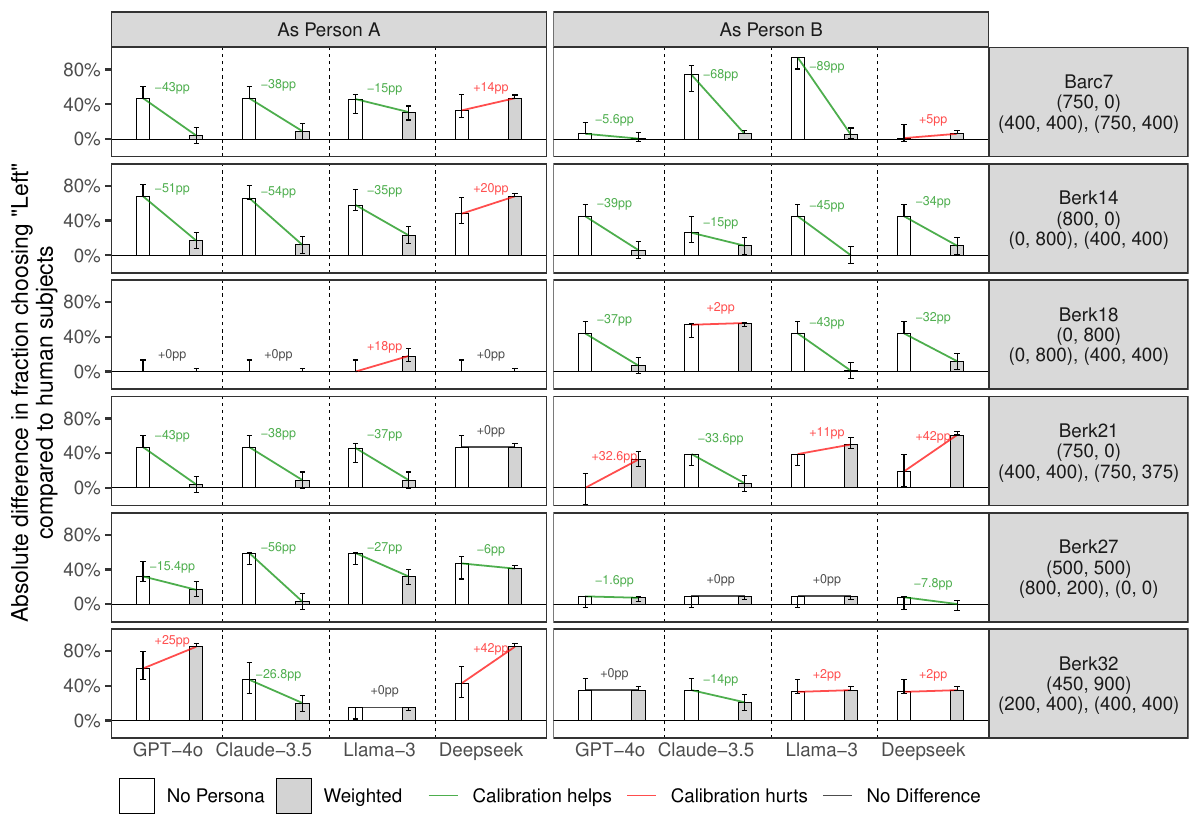}
\end{minipage}
\end{center}
\begin{footnotesize}
\begin{singlespace}
\vspace*{-0.12in}
\emph{Notes:}
This figure compares the performance of persona-less and calibrated samples of \aisub{s} using sequential games from \citeauthor{charness2002understanding}.
Rows correspond to different game scenarios, and columns to different roles (Person A or Person B).
The x-axis shows different LLMs, and the y-axis shows the absolute difference between the fraction of AI agents and the fraction of human subjects choosing ``Left'' in the original experiment.
White bars correspond to the persona-less \aisub{s}, and the grey bars correspond to the calibrated \aisub{s}.
Error bars show 95\% Wilson confidence intervals for the AI proportion, symmetrically shifted to the absolute difference from the human proportion.
See \url{www.expectedparrot.com/content/9ff43614-2726-440b-a745-f325555eab4b} for simulation code.
\end{singlespace}
\end{footnotesize}
\end{figure}

Figure~\ref{fig:charness_rabin_mixture} plots the results of the experiment, and its notes contain a linked notebook with the code for the simulations.
The y-axis is the absolute difference between the fraction of AI agents and humans choosing ``Left.'' 
For each model, the white bar shows the responses of the persona-less \aisub{s}, and the grey bar shows the responses of the calibrated sample.
The differences are substantial: across all models, the responses of the weighted \aisub{} samples are much closer to those of the human subjects.
The MSE of the calibrated samples ($0.094$) is about half of the persona-less control MSE ($0.182$).
This improvement suggests that theory-grounded persona mixtures can indeed capture meaningful patterns in human decision-making that transfer across game formats.

These results show that we are not limited to whether off-the-shelf \aisub{s} respond like humans.
Instead, we can construct \aisub{s} that respond more like humans in particular settings using personas based on some relevant theory.
We can then use these calibrated agents to explore new scenarios plausibly governed by the same theory with more confidence that their responses will predict the relevant human responses.
Indeed, \citeauthor{manning2025general} show that by constructing and then validating agents in this way (on some of these very games), we can substantially improve their ability to predict human responses in novel settings where no prior human data exists.

\subsection{\emph{Status quo} bias in decision-making~\citep{samuelson1988status}} 
\label{sec:samuelson_zeckhauser}
\cite{samuelson1988status} demonstrated in several decision-making scenarios that people are more likely to select an option if it is presented as the status quo.
In one scenario, they asked subjects to allocate a safety budget between automobiles and highways.
The instructions were:
\begin{quote}
``The National Highway Safety Commission is deciding how to allocate its budget between two safety research programs: 
(i) improving automobile safety (bumpers, body, gas tank configurations, seatbelts), and 
(ii) improving the safety of interstate highways (guard rails, grading, highway interchanges, and implementing selectively reduced speed limits).''
\end{quote}
Subjects were then asked to choose between four funding allocations: ($70\%$ auto, $30\%$ highways), ($60\%,40\%$), ($50\%,50\%$), or ($30\%,70\%$).
The main experimental manipulation was to present the options relative to different status quo allocations.\footnote{In the $(60,40)$ status quo framing, the options were: ``Decrease the highway program by 10\% of budget and raise the auto program by a like amount $(70,30)$, maintain present budget amounts for the programs $(60,40)$...''}

We recapitulate \citeauthor{samuelson1988status}'s experiment.
First, we endow each \aisub{} with one of twelve beliefs about the importance of car and highway safety.
For each belief, we create a ``car owner'' and a ``non-car owner'' \aisub{}.
We then have each of the 24 agents choose one of four funding allocations in four scenarios: each one phrased to indicate a different option as the \emph{status quo}.
We run this experiment across all five LLMs with temperature set to 1, generating multiple choice occasions per agent-scenario combination.
We use \textsc{Gpt-4}, \textsc{Gpt-4o}, \textsc{Claude-Sonnet-3.5}, \textsc{Claude-Sonnet-3.7}, and \textsc{Deepseek}.
Using this combination of models allows us to compare the \aisub{s}' responses as the capabilities of LLMs progress, as well as between different LLM providers.
We provide further simulation details inside the notebook linked in the Table~\ref{tab:zeckhauser} notes.

It is worth noting that this within-subject experimental design is possible because \aisub{s} have no memory unless we explicitly provide them with information about their previous decisions. 
This would be impossible with human subjects.
They would likely become aware of the experimental manipulation after exposure to each status quo framing.

We model the \aisub{s}' responses as discrete choices over the four allocations using a conditional logit specification and report the results in~Table~\ref{tab:zeckhauser}.
We pool data across all five LLMs but allow coefficients to vary by model using model-specific interactions.
Each observation is one choice occasion (an agent choosing among four alternatives in a given scenario), and the model estimates the probability of choosing each alternative conditional on the choice set.
\begin{table}[h!] 
\begin{center}
\begin{minipage}{1 \linewidth}                          
\caption{Recapitulation of \cite{samuelson1988status} with \aisub{s}}
\label{tab:zeckhauser}     
\centering
\footnotesize
\setlength{\tabcolsep}{-4pt}
\begin{tabular}{@{\extracolsep{.25pt}}lD{.}{.}{-3} D{.}{.}{-3} D{.}{.}{-3} } 
\\[-1.8ex]\hline 
\hline \\[-1.8ex] 
 & \multicolumn{3}{c}{\textit{Dependent variable:}} \\ 
\cline{2-4} 
\\[-1.8ex] & \multicolumn{3}{c}{Choice of Allocation} \\ 
\\[-1.8ex] & \multicolumn{1}{c}{(1)} & \multicolumn{1}{c}{(2)} & \multicolumn{1}{c}{(3)}\\ 
\hline \\[-1.8ex] 
 Claude-3.5: AutoShare & -0.010^{*}$ $(0.002) & -0.011^{*}$ $(0.002) & -0.015^{*}$ $(0.002) \\ 
  Claude-3.7: AutoShare & -0.004^{*}$ $(0.002) & -0.005^{*}$ $(0.002) & -0.007^{*}$ $(0.002) \\ 
  Deepseek: AutoShare & 0.017^{*}$ $(0.002) & 0.018^{*}$ $(0.002) & 0.017^{*}$ $(0.002) \\ 
  GPT-4: AutoShare & -0.008^{*}$ $(0.002) & -0.009^{*}$ $(0.002) & -0.014^{*}$ $(0.002) \\ 
  GPT-4o: AutoShare & -0.004^{*}$ $(0.002) & -0.004^{*}$ $(0.002) & -0.008^{*}$ $(0.002) \\ 
  Claude-3.5: I(StatusQuo) &  & 0.911^{*}$ $(0.054) &  \\ 
  Claude-3.7: I(StatusQuo) &  & 0.703^{*}$ $(0.055) &  \\ 
  Deepseek: I(StatusQuo) &  & 0.762^{*}$ $(0.055) &  \\ 
  GPT-4: I(StatusQuo) &  & 0.931^{*}$ $(0.054) &  \\ 
  GPT-4o: I(StatusQuo) &  & 0.316^{*}$ $(0.058) &  \\ 
  Claude-3.5: $|$DistStatusQuo$|$ &  &  & -0.032^{*}$ $(0.002) \\ 
  Claude-3.7: $|$DistStatusQuo$|$ &  &  & -0.020^{*}$ $(0.002) \\ 
  Deepseek: $|$DistStatusQuo$|$ &  &  & -0.029^{*}$ $(0.002) \\ 
  GPT-4: $|$DistStatusQuo$|$ &  &  & -0.040^{*}$ $(0.002) \\ 
  GPT-4o: $|$DistStatusQuo$|$ &  &  & -0.023^{*}$ $(0.002) \\ 
 \hline \\[-1.8ex] 
Observations & \multicolumn{1}{c}{6,999} & \multicolumn{1}{c}{6,999} & \multicolumn{1}{c}{6,999} \\ 
Log Likelihood & \multicolumn{1}{c}{-9,630.112} & \multicolumn{1}{c}{-9,173.293} & \multicolumn{1}{c}{-9,188.728} \\ 
\hline 
\hline \\[-1.8ex] 
\end{tabular}

\end{minipage}
\end{center}
\begin{footnotesize}
\begin{singlespace}
\emph{Notes:}
This table reports the results from our replication of \cite{samuelson1988status} with \textsc{Gpt-4}, \textsc{Gpt-4o}, \textsc{Claude-Sonnet-3.5}, \textsc{Claude-Sonnet-3.7}, and \textsc{Deepseek}.
Columns show logit discrete choice models estimating the probability of choosing each of the four allocations.
Observations represent one funding allocation choice---excluding one observation with missing data.
Each choice occasion is expanded to four rows (one per alternative), with the dependent variable indicating the selected option.
We pool data across all five LLMs but estimate model-specific coefficients through interactions with model indicators.
Independent variables are the proportion of funds allocated to automobile safety for each option, an indicator of whether or not the option was framed as the status quo, and the absolute distance between each option and the status quo.
See \url{www.expectedparrot.com/content/0bacba4d-91fa-4005-8f7d-abd34a4a1f52} for prompts and code.
\starlanguage{}
\end{singlespace}
\end{footnotesize}
\end{table}

In Column (1), we see that \aisub{s}, with the exception of \textsc{Deepseek}, are less likely to choose an alternative as more funds are allocated to highway safety (AutoShare).
In Column (2), we add a variable indicating whether an allocation was framed as the status quo.
All \aisub{s} exhibit considerable status quo bias, although \textsc{Gpt-4o} does so to a lesser extent.
In Column (3), we replace the indicator variable with a variable measuring the absolute distance between each allocation's auto share and the status quo allocation's auto share.
We see that allocations ``farther'' from the status quo are, on average, less likely to be chosen for all models.
Both specifications (Columns 2 and 3) show strong evidence of status quo bias, with Column (2) fitting the data slightly better.
Overall, our results suggest that \aisub{s} shift their preferences toward the status quo allocation---similar to human subjects. 
Even with substantial post-training fine-tuning, possibly suboptimal heuristic features of human choice remain relevant to LLM behavior.

\subsection{A complexity view of prospect theory~\citep{Oprea2024lottery}} 
\label{sec:oprea}
Prospect theory predicts that when making risky choices, people overweight small probabilities, underweight large probabilities, are risk-averse over gains, and risk-seeking over losses~\citep{kahneman1979prospect}.
This gives rise to the classic ``fourfold pattern'' of risk attitudes.
\cite{Oprea2024lottery} examined if this fourfold pattern can emerge in the absence of uncertainty.
In a novel design, participants were given endowments and faced sets of 100 closed boxes, each containing either a gain or a loss with probability $p \in \{0.1,0.25,0.75,0.9\}$.
In the ``gain'' framing, some boxes contained \$25 and the rest \$0; in the ``loss'' framing, some contained $-\$25$ and the rest \$0.

In the ``lottery'' condition, for each probability $p$ and each framing (gain or loss), certainty equivalents were elicited by asking participants to choose between a sure amount and the outcome from one randomly selected box.
This lottery condition is equivalent to a standard binary-choice lottery.
As expected, participants' certainty equivalents collectively displayed the fourfold pattern.

A second, ``mirror'' condition used the same box setup, probabilities, framings, and dollar amounts, but altered the payoff rule.
Instead of randomly choosing one box, all boxes were opened and participants received the average payoff across them.
Unlike the lottery condition, this decision involves no risk: the payoff is deterministic and always equals the expected value ($\pm 25 \times p$).
The rational certainty equivalent is therefore the expected value.
The subjects' responses to risk should not matter, and the fourfold pattern should not appear.

\citeauthor{Oprea2024lottery} finds that participant responses in the mirror condition still displayed the pattern in the aggregate.
This suggests the pattern does not arise solely from probability weighting or reference dependence, and that prospect theory is not just a theory of uncertainty. 
Instead, \citeauthor{Oprea2024lottery} argues it reflects systematic responses to complexity under cognitive constraints.
These findings have sparked some debate. 
\cite{UriCritique2025Oprea} contend that the fourfold pattern appears in the mirror condition because subjects are confused about the instructions, and interpret the condition as a risky lottery---likely because of prior exposure to conventional certainty equivalent tasks.

\aisub{s} may offer a unique perspective to help us understand features of this debate.
In simulations, we can systematically vary features of their personas that might indicate which perspective is correct.
This would otherwise be difficult, if not impossible, to do with human subjects.

We extend \citeauthor{Oprea2024lottery}'s central experiment using \aisub{s}.
We endow each agent with $\$50$, and then ask it to provide certainty equivalents for 100 box sets with probabilities $p \in \{0.05, 0.10, \dots, 0.95\}$, adding 14 probability conditions beyond those in the original experiment for each condition-framing combination. 
We construct three personas with the goal of manipulating the LLMs as if they had varying levels of available cognitive resources---a well-established method for identifying complexity effects \citep{Oprea2024Complexity}. 
The personas are: (i) ``you are very bad at math,'' (ii) ``you are average at math,'' and (iii) ``you are a mathematical genius.''
If the fourfold pattern reflects behavior under complexity rather than under risk, not only should the lottery and mirror conditions be similar, but we should also expect larger deviations from expected utility---toward prospect-theoretic predictions---among the lower-math ability personas.
We run simulations for each agent, including a baseline agent without any persona, 10 times on five LLMs (\textsc{Gpt-3.5-Turbo}, \textsc{Gpt-4o}, \textsc{Claude-Haiku-3}, \textsc{Claude-Sonnet-3.5}, and \textsc{Deepseek}), all at temperature~1.
This generates 15,200 observations.\footnote{4 agents $\times$ 2 conditions $\times$ 2 framings $\times$ 19 probabilities $\times$ 5 models $\times$ 10 repetitions = 15{,}200.}

\begin{figure}[h!]
\begin{center}
\begin{minipage}{1 \linewidth}
\caption{Recapitulation and extension of \cite{Oprea2024lottery}} 
\vspace*{-0.1in}
\label{fig:oprea}
\includegraphics[width = \linewidth]{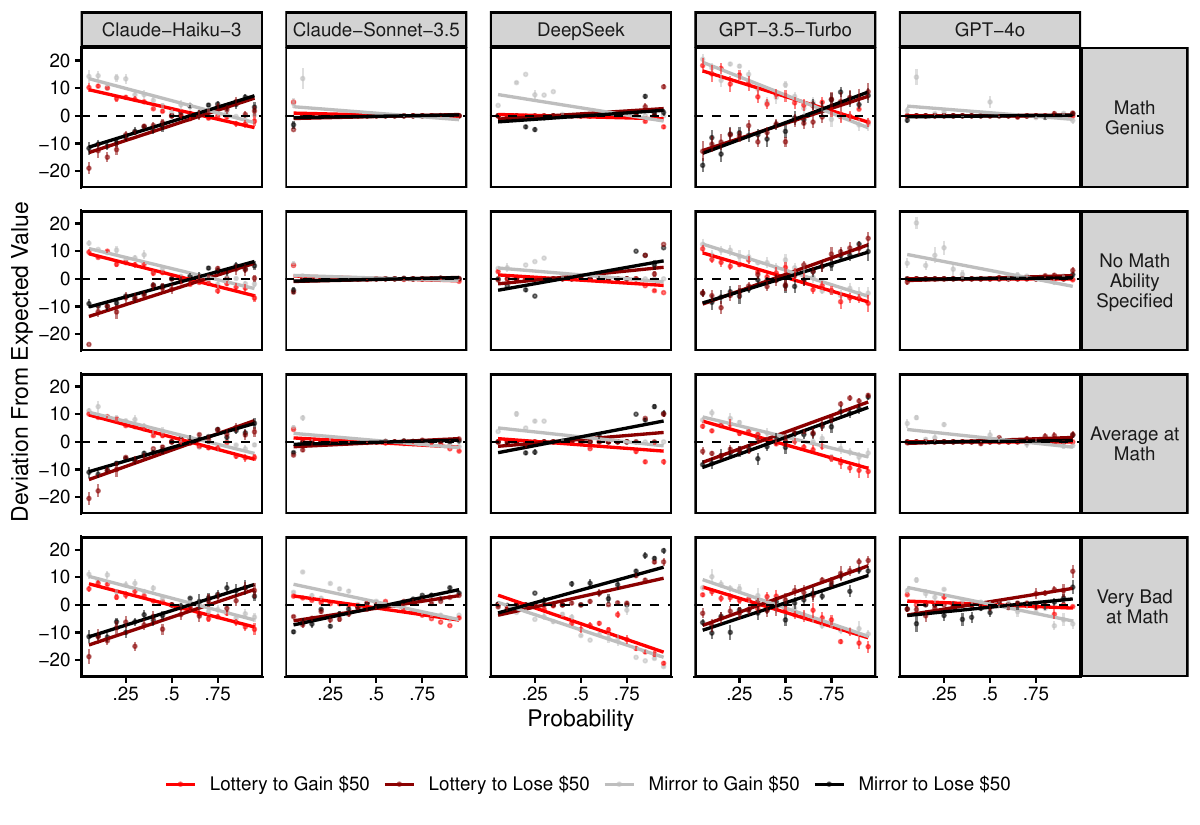}
\end{minipage}
\end{center}
\begin{footnotesize}
\begin{singlespace}
\vspace*{-0.15in}
\emph{Notes:}
This figure shows the results of our recapitulation and extension of \cite{Oprea2024lottery} with \aisub{s}. 
The x-axis is the probability of winning or losing \$50, and the y-axis is the deviation from the lottery's (or mirror's) expected value. 
Columns correspond to different LLMs, and rows to whether different agents were endowed with different mathematical abilities.
Each dot is the mean certainty equivalent across \aisub{s} (with 95\% confidence intervals), and the solid lines are OLS fits within each framing-condition combination.
Colors represent condition-framing combinations.
See \url{www.expectedparrot.com/content/05009e7a-84e8-4c14-bc5d-de0408df89a2} for simulation code.
\end{singlespace}
\end{footnotesize}
\end{figure}

Figure~\ref{fig:oprea} displays the simulation results.
The x-axis shows the probability of winning (or losing) \$50, and the y-axis shows the deviation of each agent's certainty equivalent from the lottery's (or mirror's) expected value.
Columns correspond to different models, and rows indicate the \aisub{}'s assigned mathematical ability.
Colors represent condition-framing combinations.
Each point is the mean certainty equivalent across 10 runs of a persona at a given probability, with 95\% confidence intervals.
Lines show OLS fits regressing deviation from expected value on probability for each condition independently for losses and gains.

Across models, the lottery and mirror conditions generally move in lockstep: either both show the characteristic X-shape or both are flat at the expected value.
There are no instances where the lottery displays the fourfold pattern but the mirror does not.
Notably, there are also no panels where the slopes are reversed: under gains, all regression lines are weakly decreasing in $p$, and under losses, all regression lines are weakly increasing in $p$.
Less advanced models---\textsc{Gpt-3.5-Turbo} and \textsc{Claude-Haiku-3}---show pronounced deviations between small and large probabilities.
By contrast, more capable models---\textsc{Gpt-4o}, \textsc{Claude-Sonnet-3.5}, and \textsc{Deepseek}---more often select expected values as a baseline, particularly when endowed with stronger mathematical abilities, but most clearly exhibit the fourfold pattern when prompted as ``very bad at math.''

On its face, our recapitulation supports \citeauthor{Oprea2024lottery}'s claim that the fourfold pattern is not a special phenomenon of risk.
However, it is worth considering systematically the plausible explanations for these AI responses---both for exposition of \emph{Homo silicus} and as commentary on the results.

Could our findings reflect mere noise or rote memorization by the LLMs? 
Pure noise seems implausible, given the consistency of the fourfold pattern across five separate models and thousands of simulations. 
We would expect to see at least one regression line with an inverted slope relative to the prospect-theoretic predictions if this were the case.
Memorization of \citeauthor{Oprea2024lottery} is also unlikely. 
Early drafts may be in training data, but the paper only gained broad visibility in 2024, and its deterministic mirror condition is novel.
We also vary probabilities and payoffs beyond the original, so a narrow script tailored to those exact experiments would not suffice.
It would represent a remarkable degree of transfer learning.

Another possible explanation is that LLMs may have ingested decades of lottery literature, so they default to a standard prospect-theoretic response. 
This is similar to \citeauthor{UriCritique2025Oprea}'s argument that \citeauthor{Oprea2024lottery}'s subjects may have reverted to familiar responses because they misunderstood the payoff structure.
The variation in responses across personas suggests that the ``confusion with lotteries'' perspective is unlikely.
The more capable models largely select the expected value in all conditions when assigned a high, medium, or unspecified mathematical ability, but generate responses corresponding to an increasingly pronounced fourfold pattern when they are ``very bad at math.''
All three of the more capable LLMs---each independently developed with its own training data and fine-tuning processes---appear to have learned some relationship between stated mathematical ability and deviation away from expected utility theory in a way that aligns precisely with prospect theory.
After an extensive search of the literature, we found no such robustly reported relationship for an LLM to have learned to apply by default.\footnote{While there is some evidence of a weak negative correlation between cognitive ability and risk aversion in gains, it is not large, and we did not find published evidence for a general relationship relating to relative losses \citep{Lilleholt2019RiskCog,Dohmen2010RiskCog}.}

The most plausible explanation in our view is that LLMs, having ingested innumerable examples of humans solving problems and demonstrating their preferences over possible states of the world---one of the implicit sources of latent social information that we discussed in the introduction---learned systematic relationships about how humans trade off features of decision-making problems under complexity.
This view is consistent with both the similarity between lottery and mirror conditions, and the increasingly prospect-theoretic responses as the agents get worse at math.
Of course, this does not provide definitive evidence on whether \citeauthor{Oprea2024lottery} or \citeauthor{UriCritique2025Oprea} are correct.
For now, real human data are necessary to shed light on such a question.
But these findings highlight the power of using LLMs to study subtle behavioral patterns likely present in their training corpora.

\subsection{Labor-labor substitution under a minimum wage~\citep{laborHorton2025}} 
\label{sec:horton}
\cite{laborHorton2025} reports the results of an experiment where employers were randomly assigned minimum wages in an online labor market: when applying for a job, applicants had to bid up to meet the employer's randomly assigned minimum wage.
A key finding of this experiment was that there was little reduction in hiring but a substantial shift towards more productive workers, as proxied by past worker earnings and experience. 
The possibility of this labor-labor substitution margin had been noted in the literature, but had been difficult to pin down empirically in less controlled settings.

We explore the labor-labor substitution margin with \aisub{s}.
We create scenarios where we ask agents (employers) to select from pairs of applicants that vary in their work experience and requested wages.  
In our scenario, employers are hiring for the role of dishwasher, and we inform them of the typical wage for this role.
The scenario in the prompt is:
\begin{quote}
You are hiring for the role of “Dishwasher.”
You have 2 candidates.\\
Person 1: Has 1 year(s) of experience in this role. Requests \$[\texttt{wage\_ask\_1}]/hour.\\
Person 2: Has 0 year(s) of experience in this role. Requests \$[\texttt{wage\_ask\_2}]/hour.\\
Who are you hiring? You must fill this role.
\end{quote}
In our setup, Person 1 is the ``experienced'' worker and Person 2 is the ``inexperienced'' worker.
The inexperienced worker always asks for \$13/hour, and the experienced worker asks for a wage between \$12 and \$17/hour.
The employer is unknowingly and randomly assigned to either no minimum wage or a minimum wage of \$15/hour.
With a minimum wage, wage asks below the minimum wage threshold are forced up to \$15/hour for both workers.

We repeat our experiment twice: once when we tell the employer that the typical wage ask for the position is \$12/hour, and once when we provide no information about the typical wage.
We elicit the \aisub{s}' hiring decisions in 24 scenarios with 43 different models five times each at a temperature of 0.5.\footnote{There are 6 possible wage asks, 2 minimum wage conditions, and 2 reference wage conditions. $6 \times 2 \times 2 = 24$.} 
We ask the employer to always hire a candidate.
Our empirical approach is to regress the hired worker's wage and experience on an indicator for the minimum wage treatment status.
\begin{table}[h!] 
\begin{center}
\begin{minipage}{1 \linewidth}                          
\caption{Effects of minimum wage on observed wage and hired worker attributes}  
\label{tab:horton}     
\centering                                                        
\small
\begin{tabular}{@{\extracolsep{15pt}}lcccc} 
\\[-1.8ex]\hline 
\hline \\[-1.8ex] 
 & \multicolumn{4}{c}{\textit{Dependent variable:}} \\ 
\cline{2-5} 
\\[-1.8ex] & \multicolumn{2}{c}{Hire wage} & \multicolumn{2}{c}{Hire experience} \\ 
\\[-1.8ex] & (1) & (2) & (3) & (4)\\ 
\hline \\[-1.8ex] 
 I(Minimum Wage) & 1.314$^{*}$ & 1.984$^{*}$ & 0.081$^{*}$ & 0.265$^{*}$ \\ 
  & (0.067) & (0.053) & (0.018) & (0.023) \\ 
  & & & & \\ 
\hline \\[-1.8ex] 
Reference Wage & No & Yes & No & Yes \\ 
Model FE & Yes & Yes & Yes & Yes \\ 
Observations & 2,580 & 2,579 & 2,580 & 2,579 \\ 
R$^{2}$ & 0.273 & 0.628 & 0.226 & 0.179 \\ 
\hline 
\hline \\[-1.8ex] 
\end{tabular}

\end{minipage}
\end{center}
\begin{footnotesize}
\begin{singlespace}
\emph{Notes:} This table shows regressions where the independent variable indicates whether a minimum wage was imposed on the AI employer.
Dependent variables are (1) and (2) the hired worker's hourly wage and
(3) and (4) their years of experience.
Columns (1) and (3) show results without a reference wage, and Columns (2) and (4) with a reference wage.
Employer-level observations are sampled from 43 different models.
See \url{www.expectedparrot.com/content/15eb59ad-f7c8-4c9b-87d1-a58cfecc13ec} for prompt details and the list of models.
\starlanguage{}
\end{singlespace}
\end{footnotesize}
\end{table}
Table~\ref{tab:horton} reports the regression estimates; the notes include a linked notebook with the list of models and all simulation details.
All specifications include model fixed effects, and standard errors are clustered at the model level.
Column~(1) shows that without a reference wage, the minimum wage caused an increase of about \$1.3/hour in the wage of the hired worker.
Column~(2) shows that when a reference wage is provided, this effect grows by nearly 50\%.
Positive coefficients are expected because hiring was mandatory for the AI employer.

When we regress the hired worker's experience on the presence of a minimum wage, we see similarly substantial effects.
Column~(3) shows that without a reference wage, the minimum wage increased the hired worker's years of experience by about a month.
With a reference wage, Column~(4) shows that the introduction of the minimum wage caused the employer to hire workers with roughly three more months of experience on average.
We return to why the reference wage matters in Section~\ref{sec:conceptual_issues}, but even across simulations of dozens of independently developed models, these increases in wage and experience are directionally consistent with \citeauthor{laborHorton2025}'s findings.

\section{Why and when we can learn from \emph{Homo silicus}} 
\label{sec:when-why}
In the previous section, we demonstrated that \aisub{s} can simulate features of human behavior in experiments along with various methods to facilitate such simulations.
Next, we examine why LLMs possess these capabilities in the first place, which can inform when the approach is most likely to be useful.
We begin by describing the data that enables highly general AI simulations, and then discuss how economists can conceptualize them in their methodological toolkit.

\subsection{Why LLMs can simulate human behavior}
Good predictions start with good training data.
LLMs can flexibly predict human responses because their training data contain rich information about the social world, and their architectures and post-training processes allow them to learn and apply generalizable human behavioral patterns from that data.

Figure~\ref{fig:diagram_knowledge} depicts the data pipeline from the social world to LLM predictions about the social world.
LLMs gather informative signals about the social world through two distinct paths: 
(i) explicitly, by memorizing findings and theories from the social science literature and 
(ii) implicitly, by ingesting text that captures aspects of the social world, such as news articles, social media content, books, and so on.

\begin{figure}[h!]
\begin{center}
\caption{Why LLMs are good predictive models of the social world} \label{fig:diagram_knowledge}
\begin{tikzpicture}[%
>=Latex,
node distance = 1.3cm and 4.5cm,
box/.style  = {draw, rounded corners, inner sep=6pt, text width=3.3cm, align=center, font=\footnotesize}
]

\node[box] (world) {Social world};

\node[box, right=of world, text width=4.2cm] (media)
{Non-scholarly human\\texts about social world};

\node[box, below=1.6cm of world] (liter)
{Social science\\literature};

\node[box, below=1.6cm of media] (train) {Training data};

\node[box, below=1.6cm of train]  (fitted) {Fitted model\\(LLM)};

\node[box, below=1.6cm of liter] (pred) {Predictions about \\the social world};

\draw[->] (world) -- (media)
node[midway, above, sloped, font=\itshape\scriptsize, align=center]
{\textbf{(ii)} Everything written/recorded};

\draw[->] (world) -- (liter)
node[midway, right, font=\itshape\scriptsize, align=center]
{\textbf{(i)} Everything that\\gets researched\\and published};

\draw[->] (media) -- (train)
node[midway, left, font=\itshape\scriptsize, align=center]
{Some unclear\\selection process};

\draw[->] (liter) -- (train)
node[midway, above, sloped, font=\itshape\scriptsize, align=center]
{Some unclear selection process};

\draw[->] (train) -- (fitted)
node[midway, left, font=\itshape\scriptsize, align=center]
{Neural network\\architecture\\and post-training};

\draw[->] (fitted) -- (pred)
node[midway, above, font=\itshape\scriptsize, align=center]
{Prompting};

\end{tikzpicture}
\end{center}
\begin{footnotesize}
\begin{singlespace}
\vspace*{-0.05in}
\emph{Notes:} This figure is a simplified depiction of the data pipeline from the social world to the LLM's training data, and then to the LLM's predictions about the social world. 
It highlights that LLM simulations can match human behavior in the real world through two distinct paths: (i) by acting in accordance with explicit theories and findings in the social science literature, or (ii) implicitly by ingesting elements of human behavior ``in action'' captured in non-scholarly human-generated text.
\end{singlespace}
\end{footnotesize}
\end{figure}

The social science literature path contains records of generalizable patterns, processes, relationships, and theories about human behavior---precisely the kinds of insights we would like LLMs to internalize and apply more broadly.
Consider that at the core of economic analysis is the idea that humans are goal-oriented agents maximizing some objective under constraints \citep{vNM1944}, or the empirical evidence that humans often derive utility with diminishing marginal returns to consumption \citep{samuelson1937} and exhibit strong social preferences by valuing fairness, reciprocity, and others' welfare alongside their own \citep{fehrSchmidt1999}.
These concepts do not just magically exist in the ether.
They have been codified through decades of research, dramatically improving economists' understanding of human behavior.
These are the essential ideas that any ``general'' human model should take into account.
And while learning from the social science literature comes with distinct challenges, which we discuss in more detail in Section~\ref{sec:conceptual_issues}, it is a necessary ingredient for LLMs to effectively simulate human responses in many settings.

In contrast to the social science literature path, the non-scholarly, human-generated text path provides LLMs with true aspects of the social world ``in action:'' how people communicate, make decisions, interact with each other, and perceive the world. 
We might be skeptical that a language model could, say, represent physics we have yet to theorize, but it surely has internalized relevant knowledge about the social world that has never been written down in academic work.
This is not to say that the data are the social world or a perfect representation of it.
Rather, it contains countless examples that reiterate basic economic relationships across a wide range of contexts.

Consider the observation that people prefer lower prices and greater quantities, all else equal. 
Any economic textbook would contain this observation either in some written form or with the condition $\frac{\partial}{\partial p}\bigl(u(x^*) - px^*\bigr) < 0$. 
But the same observation also appears in widely-read, pre-social science sources.
In the Book of Amos, the prophet denounces unfair traders: ``\emph{And the Sabbath, that we may offer wheat for sale, that we may make the ephah small} [a weight to put on a balance] \emph{and the shekel great and deal deceitfully with false balances.}''

What about more subtle knowledge, such as the literature on cheap talk in bargaining games? 
There, we would not expect the LLM to learn every detail and figure out all implications of \cite{crawford1982strategic}, but its data contain sources that capture some of the basic ideas and tensions.
For example, ``\emph{`It's no good, it's no good!' says the buyer---then goes off and boasts about the purchase}'' (Proverbs 20:14) captures the idea that in bargaining, the buyer and seller are not always forthright.

Modern non-scholarly sources echo these dynamics as well.
Over the past three decades, an increasing share of economic activity has occurred online, creating an extensive record of transactions, reviews, negotiations, and market interactions.
These digital traces provide ample raw material from which LLMs learn realistic representations of human decision-making in many settings.

Importantly, both the implicit and explicit paths are shaped by distinct and often opaque selection mechanisms.
Peer review determines which studies enter the formal literature, and individual self-presentation---what people choose to write, post, or record---filters the broader corpus of human text.
Model developers then apply a further layer of curation when assembling the final training data, excluding or weighting sources according to proprietary criteria.

Neither stream of information is complete or representative, which is why AI simulations should not be treated as unequivocal evidence about human behavior.
Over time, however, we expect that the models most frequently used for simulation will converge toward greater transparency and openness, with clear data inclusion standards, explanations of post-training processes, and documentation.
While open models are not yet as advanced as the most capable models and not viable for many AI experiments, they are improving rapidly.
When possible, we fully endorse the use of open models.
For now, there remain trade-offs between model capability and openness, though these have narrowed considerably even since the first release of ChatGPT.

While training data are key to the ability of LLMs to simulate human behavior, these information sources alone do not create a coherent model.
LLMs' architectures and post-training processes translate rich training data into useful predictions about human behavior. 
Modern transformer LLMs are flexible sequence-to-sequence estimators: they can approximate complex mappings from contextual variables such as prices, payoffs, and information sets to a distribution of actions in their textual responses \citep{yun2019transformers}.
Their training objective rewards outputting empirically truthful conditional probabilities over their data \citep{gneiting2007strictly}---minimizing log loss is a strictly proper scoring rule.

Post-training methods (like instruction tuning or reinforcement learning with human feedback) steer models away from brittle responses driven by spurious patterns in their training data and toward cooperative, instruction-following behavior---we exploit this property when we endow agents with roles and preferences.
The most advanced frontier models even have specific reasoning capabilities which allow them to better ``think through'' how a human would respond based on a given set of characteristics.
Together, massive training corpora combined with these architectural and training innovations enable LLMs to capture generalizable patterns of human behavior that can be used to generate predictions in a wide range of settings.

\subsection{\emph{Homo silicus} as theory}
\label{sec:theory}
In Section~\ref{sec:experiments}, we used elements of the research processes traditionally associated with empirical work to construct and analyze our AI simulations.
We designed experiments, collected data, and used statistical analyses to summarize the results.
Despite these ostensible similarities to applied work, many of the most valuable use cases and insights gained are more akin to the practice of economic theory.

To see why, consider an economist who is trying to understand the following puzzle: why do large firms pay higher wages than observationally equivalent small firms? 
The researcher begins by trying to create a model where such a gap exists in equilibrium. 
If she assumes perfectly competitive labor markets and costless mobility, her model would likely predict no firm-size wage differences---workers are simply paid their marginal product. 
Of course, this disappointing first step does not prove that theorizing is worthless. 
Instead, it highlights that the baseline model omits important features of real labor markets. 

The next step in the researcher's journey is to create richer models.
With each new theory direction, she would attempt to formalize a different real-world mechanism. 
At each step, the researcher must draw on domain knowledge and the literature to specify a new model, work out the equilibrium (often painfully), and derive useful, testable predictions. 

With \emph{Homo silicus}, we can easily construct a large number of AI simulations---based on a world model informed by empirical data---that might bear on the firm-size premium.
For example, suppose we simulate a job seeker weighing offers from a small startup and a large incumbent. 
If an AI agent spontaneously emphasizes differences in layoff risk and promotion timelines---and accepts a lower starting wage at the larger firm because it anticipates faster wage growth and insurance value---this suggests a mechanism and maybe even an empirical design: measure expected wage paths and mobility frictions by firm size. 

Similar to modeling, simulations can lead us in the wrong direction. 
But they likely make expensive data collection more useful by helping us target efforts where the marginal value of information is highest.
Economists have long been aware of how seemingly simple micro-decision making can lead to surprising and subtle equilibrium implications \citep{AkerlofYellen1985}.
It is not hard to imagine how the ability to easily conduct rich, complex simulations might lead to surprising and generative insights.
Indeed, this is one of the major differences between the physical and social sciences: the ability to iterate and run follow-up experiments.
Given the differences in predictive power between physical and social scientific theories, there is reason to believe a low-cost and high-speed approach to ``learning by doing'' \citep{Arrow1962} at speeds closer to the lab bench could be a powerful tool for progress.

While the many parallels with economic theory are clear, a key difference---and an advantage---of \emph{Homo silicus} is its flexibility.
We can use AI simulations to generate predictions in virtually any setting described in natural language.
This is often not possible with conventional economic theory because it generally requires fixed input parameters and structural assumptions.

To make concrete what we mean by flexibility, and why economic models are relatively inflexible, consider the model in \citeauthor{charness2002understanding} used to estimate the impact of social preferences on participant allocation decisions.
For the unilateral dictator games we studied in Section~\ref{sec:experiments}, they model the utility player B derives from a given allocation $\bm{\pi} = (\pi_A, \pi_B)$ as:
\begin{equation}\label{eq:cr}
U_B (\pi_A, \pi_B) = (\rho \cdot r + \sigma \cdot s) \cdot \pi_A + (1 - \rho \cdot r - \sigma \cdot s) \cdot \pi_B,
\end{equation}
where $r = 1$ if $\pi_B \ge \pi_A$ and $r = 0$ otherwise; $s = 1$ if $\pi_B \le \pi_A$ and $s = 0$ otherwise.
Here $\pi_A$ and $\pi_B$ are the payouts for players A and B in allocation $\bm{\pi}$, respectively.
$\rho$ and $\sigma$ are free parameters representing distributional preferences which \citeauthor{charness2002understanding} estimate from their human data.

With values for the free parameters, Equation~\ref{eq:cr} could be used to predict dictator preferences over a range of two-player allocations.
However, this model is fundamentally limited in two ways.
First, it cannot incorporate any additional information from a given setting that may be relevant to the dictator's decision.
Suppose that after running their experiment, \citeauthor{charness2002understanding} learn that about half of the participants are friends and believe this might strongly interact with baseline social preferences.
To use this information to improve the model's predictions, say, as an extra variable, one must make additional assumptions and re-estimate all parameter values.
Second, this model cannot make predictions in structurally distinct games---even if these games share similar underlying motivations.
Equation~\ref{eq:cr} is useless for predicting the dictator's choice over allocations for any number of players greater than two.

Traditional machine learning models, which have become increasingly relevant in predicting human behavior in economics \citep{FudenbergLiang2019PredictPlay,peterson2021lottery,hirasawa2022using,zhu2025capturing,andrews2025transfer}, have similar limitations.
It is not possible to simply add new features to these models to make predictions in structurally distinct settings.
For example, if a regression model is trained to predict earnings linearly from education and age, but a new observation comes along that also includes cognitive ability, the linear model's estimated coefficients offer no information about the relationship between cognitive ability and earnings.
This extra column of data must either be ignored by the model or the model must be retrained from scratch using data from all three covariates.
In short, most economic and machine learning models cannot function outside the domain for which they were built.

AI simulations, however, are not subject to the same structural limitations.
In the parlance of the machine learning literature, we may think of LLMs as those models with a strong ``inductive bias'' for social science prediction problems.
We can simply adjust the setting provided in a prompt (literally rewriting the natural language description of the setting), and the LLM will predict a response based on the relationships it has learned from its training data.
While we cannot directly apply Equation~\ref{eq:cr} to three-player dictator games, we could supply the model with the parameters estimated in two-player games to an LLM and prompt it to extend the mapping in a reasonable, explicitly described way to games with any number of players.

This is possible, in part, because LLMs can reasonably impute missing information necessary to make a coherent prediction. 
Since they are next-token predictors, they can accommodate the functional form of any model written in a prompt.
Such flexibility is well-embodied in the ``P'' in ChatGPT (Generative Pre-trained Transformer)---the idea that by pretraining on a massive corpus of text, the model can generalize well.\footnote{To be sure, there are settings where parametric models are better---e.g., response data from 10,000 binary-choice lotteries \citep{peterson2021lottery} when we are specifically studying interpolation in risky choice \citep{EnkeComplexity2023,zhu2025ExplPredicting}.}

Flexibility is not without its drawbacks.
LLMs are black boxes that can fail to generalize in unexpected ways \citep{vafa2024general}.
But viewing \aisub{s} as vessels to apply interpretable economic theory to new settings is a powerful way to mitigate this risk.
Because instruction following is the central post-training objective of modern LLMs \citep{ouyang2022training,bai2022helpfulharmless}, a more reliable way to steer them is to express our assumptions as explicit instructions. 
Economic theory maps naturally into such instructions: it declares state variables and constraints, specifies an objective, and implies decision rules. 
When we write those elements clearly in the prompt (e.g., ``You only care about efficiency''), the model can apply the rule across new settings without retraining---and we can audit whether it did so. 
As we discuss in Section~\ref{sec:conceptual_issues}, explicit, unambiguous instructions are valuable because they require far less from the LLM's underlying world model.
They also make behavior checkable in simple test cases and reveal where extrapolation fails before scaling simulations up in complexity.
Put differently, economic theory supplies the instructions; LLMs supply the capacity to execute them flexibly.

We applied a version of this theory-as-flexible-instruction approach when calibrating our sample for the two-stage allocation games in Section~\ref{sec:experiments}.
For the unilateral dictator games, we endowed ``efficient'' agents to maximize total payoff, ``inequity-averse'' agents to minimize payoff differences, and ``self-interested'' agents to maximize their own payoff. 
In each of the six allocation decisions, there is a clear mathematical mapping from the instruction to the preferred allocation; only Berk26 is equivocal for the efficiency persona because both allocations are equally efficient.
After calibrating on the unilateral tasks, we applied the same agents to the two-stage games, where ``efficient,'' ``inequity-averse,'' and ``self-interested'' become more ambiguous.
For example, in Berk27, should the self-interested Player A choose ``Left'' (certain 500) or ``Right'' (either 800 or 0 depending on B's move)? Although 800 seems more likely than the spiteful 0, there is no guarantee B cooperates.
Still, the calibrated agents predicted human responses far better than persona-less LLMs, suggesting that even this imperfect theoretical mapping was relevant.

A natural question raised by these results is what they teach us about humans rather than models.
Are we actually learning that human dictators behave like mixtures of efficient, inequity-averse, and self-interested types?
More broadly, whenever an interpretable prompt improves a simulation's fit to human data, should we take that as evidence that the theoretical idea associated with that prompt is literally true of people and should anchor our economic models?
This echoes the broader idea of reverse-engineering predictive power from machine learning models as a tool for hypothesis generation \citep{MLhypothesis2024}.
For now, that inference is too strong.
LLMs remain black boxes: we do not yet know why a given prompt improves performance or how the internal representations that drive these gains relate to human behavior.
The nascent field of mechanistic interpretability is beginning to shed light on these questions \citep{olah2020circuits,elhage2021mathematical}, but it does not yet let us map prompts cleanly onto interpretable features that explain human responses.

Even so, these patterns of improved predictive accuracy provide compelling---and increasingly abundant---evidence that LLMs have internalized meaningful relationships between interpretable theory and human behavior \citep{zhu2025ExplPredicting}.\footnote{\cite{Jackson2025Mixture} offer a complementary illustration: when they optimize prompts to match human behavior across seven classic economic games, the high-fidelity prompts that emerge invoke precisely the constructs economists would anticipate---risk preferences in investment games, cooperation in public-goods games, fairness in allocation games, and so on.}
That conclusion is reinforced by the opposite pattern: prompts that endow \aisub{s} with atheoretical or scientifically meaningless traits generally do not improve predictive accuracy.
\cite{manning2025general}, for example, show that giving agents hobbies or favorite TV shows does little to improve fit on \citeauthor{charness2002understanding}'s games, either in-sample or in new settings, while theory-based social-preference personas do.
Of course, there are surely settings where the correct theory does not improve fidelity, or meaningless prompts do.
But the fact that we cannot yet make precise inferences about human behavior from which prompts improve predictive fidelity does not mean these patterns lack value.
They may still provide structured signals about which theoretical ideas are promising for modeling and exploration, even if the mechanisms underlying those gains are not yet fully understood.

\subsection{Using \emph{Homo silicus} to navigate uncharted territory}
\begin{figure}[h]
\begin{center}
\caption{How AI simulations may help or hurt social science research on new propositions}
\includegraphics[width=0.95\textwidth]{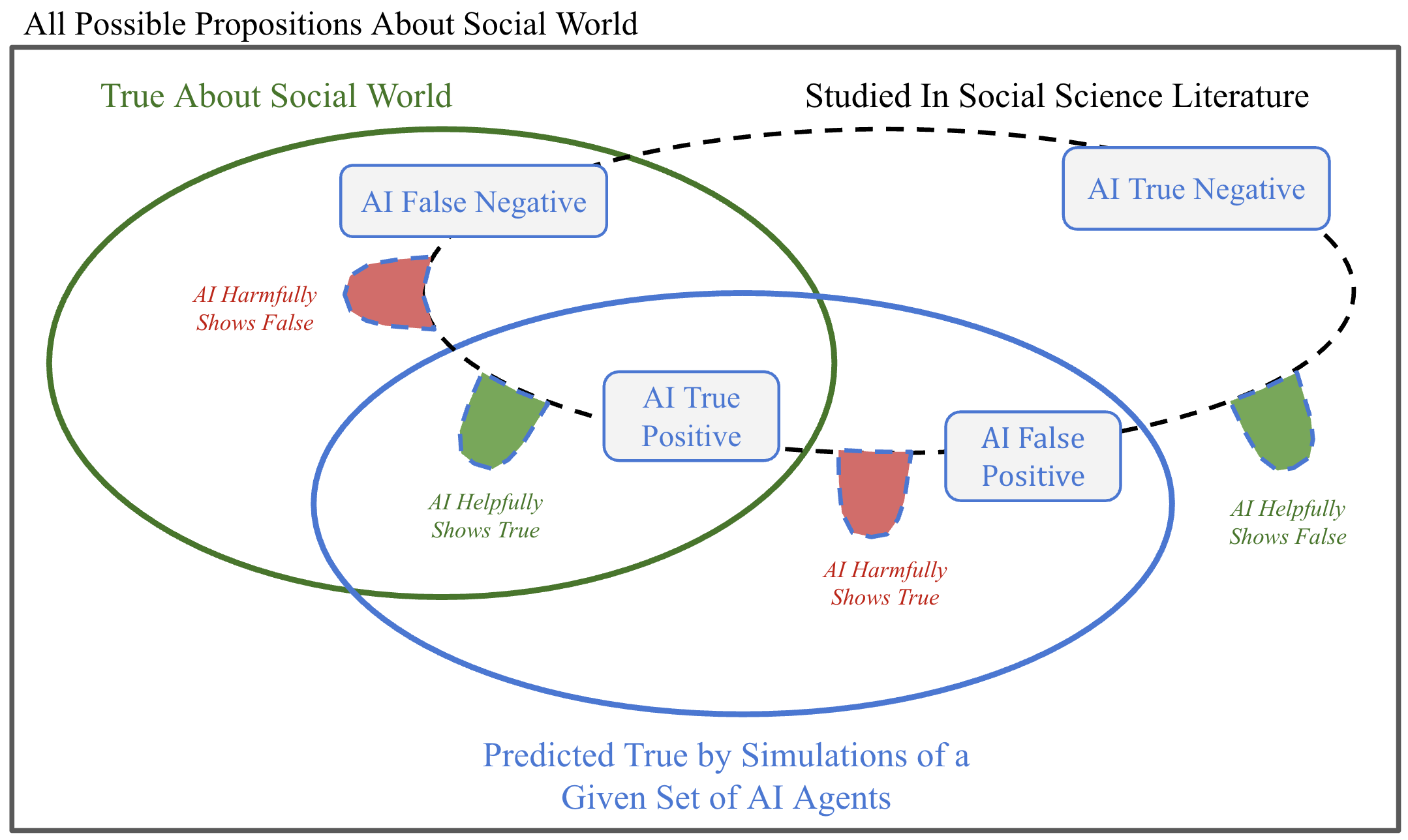}
\label{fig:diagram_ovals}
\end{center}
\begin{footnotesize}
\begin{singlespace}
\vspace*{-0.1in}
\emph{Notes:} This figure displays the predictions made by simulations of \aisub{s} (bottom center blue oval) relative to the true social world (upper left green oval).
The black dashed oval in the upper right offers the propositions that have been evaluated by social scientists---it is agnostic as to whether such propositions are recorded as true or false.
The blue ``feet'' show how AI results could expand the existing literature.
The Venn diagram implicitly offers a confusion matrix for where AI simulations can either help or hinder researchers as they explore novel ideas.
\end{singlespace}
\end{footnotesize}
\end{figure}

Figure~\ref{fig:diagram_ovals} shows how we can think about the relationship between the social world, the social science literature, and AI simulations.
The green oval (upper left) depicts the set of true propositions about the social world.
The black oval (upper right) depicts the set of propositions that have been studied and recorded in the social science literature---regardless of the findings.
These may or may not have been labeled true, which is why the oval is dashed, but they have been evaluated by researchers.
The blue oval (bottom center) depicts the propositions that simulations of a given set of \aisub{s} powered by a particular LLM would predict to be true.  
The blue ``feet'' protruding from the black oval represent how new AI results from the given agent set would expand on the existing social science literature.

Along with Figure~\ref{fig:diagram_knowledge}, this diagram shows how AI simulations show a selective and noisy slice of the social world.
They are not yet an adjudication of truth but a navigation aid: a disciplined way to mark the types of propositions that appear promising.
Read as a confusion matrix, Figure~\ref{fig:diagram_ovals} implies four possible cases:
(i) \emph{AI True Positives}: true propositions supported by AI simulations, 
(ii) \emph{AI True Negatives}: false propositions rejected by AI simulations, 
(iii) \emph{AI False Positives}: false propositions supported by AI simulations, and
(iv) \emph{AI False Negatives}: true propositions rejected by AI simulations.
Researchers stand to gain when AI simulations fall in the true positive and true negative regions. 
On the other hand, AI simulations may amplify errors when they generate false positives or neglect well-established results. 
This underscores that simulations can be either helpful when they corroborate good social science or refuse to imitate bad social science, or harmful when they introduce spurious results or fail to predict legitimate findings.

Notably, humans are not necessarily strong benchmarks for predicting the results of social science research.
A growing literature shows that even trained social scientists are surprisingly poor at forecasting the outcomes of interventions and experiments \citep{predictscience2019DellaVigna, milkmanMegastudies2021, Gandhi2025Hypothetical, Gandhi2024EffectSize, Duckworth2025ZearnMega}.
This implies that the ``bar'' for a useful predictive tool may be lower than we often assume.
AI simulations need not be perfect to be informative---they may only need to improve on human judgment.

Still, how would we know in which of the four regions we are likely to find ourselves when using AI simulations? 
Without a correct causal model, no statistical procedure can, ex ante, guarantee performance in novel settings \citep{Pearl2009Causality}.
To get unbiased estimates of economic parameters or to make predictions with guaranteed statistical bounds, one needs at least some representative data or sufficient statistics.
Only then, with the appropriate statistical frameworks, can researchers use \aisub{s}' responses as data to draw unequivocal inferences about human behavior \citep{ludwig2025llmapplied}.
In other words, researchers generally need some sort of ground truth or gold standard dataset to compare against.
In general, ground truth is the analogous human subjects' data for a given set of simulations. 

One such case where ground truth is available is when researchers have already committed to collecting a human sample and want to stretch it further.
For these settings, prediction-powered inference is especially useful.
The method starts with a ground truth sample, trains a model to predict the outcome from observed covariates, and then combines those predictions with inputs lacking outcome labels to form estimators that remain unbiased for the target parameter \citep{Angelopoulos2023Prediction, Egami2023Using}.
The more accurate the predictive model, the more precise the resulting estimates and the lower the cost of achieving a desired level of precision.
Intuitively, the machine learning model's predictions act like additional observations that are corrected using the ground truth sample.
AI simulations can be particularly effective in this role \citep{Broska2025mixed,ludwig2025llmapplied}, in no small part because of their flexibility.

Ground truth, however, is often unavailable for another one of the most useful applications of AI simulations: exploring new settings where no prior data, and therefore no guarantees, exist.
In such cases, we are not generally interested in unbiased inference, but instead in exploration and hypothesis generation---hence our continued emphasis on theory comparison.
The evidentiary burden for such research tasks is far lower than for unbiased inference.
Still, researchers should not want to explore blindly. 

The key is to construct a set of \aisub{s} that are more likely to perform well in a given setting, even if perfect fidelity cannot be guaranteed.
One way to do so, as we have shown, is to design theory-grounded agents and test them in settings where human ground truth is available and theoretically related to the target domain of interest.
If these agents can accurately predict human responses in such benchmark settings, we gain confidence that they will generalize effectively to nearby, unlabeled settings.
Conceptually, there are many distinct blue ovals in Figure~\ref{fig:diagram_ovals}, each representing predictions of simulations from a different set of \aisub{s}. 
These could be different models, prompts, or a combination of both.
The goal is to identify those that can be shifted closest to the green oval---the set of true propositions---especially for those propositions where the black oval provides reliable labels.
When the new settings lie close to these correctly labeled regions, we might expect similar predictive success---illustrated by the green ``AI helpful'' feet in the figure.

Our experiments with \citeauthor{charness2002understanding}'s unilateral and two-stage allocation games provide an example of this approach. 
We used agents grounded in social preferences to match human responses in the unilateral dictator games, then applied them to the related two-stage games where we presumed the same underlying theory (various social preferences) drove behavior. 
\cite{manning2025general} provide a full elucidation of this approach and demonstrate that the method can substantially improve the predictive power of AI simulations in entirely novel settings where no prior human data exists.\footnote{An alternative to effective prompting for improving the fidelity of simulations is to fine-tune the model on a particular dataset related to the target setting \citep{suh2025finetuningscaled,binz2025centaur}. 
However, if the goal is ideation, such fine-tuning sacrifices many of the lessons learned from identifying the appropriate theory-grounded agents in the first place.
For now, it is also far more unwieldy and can lead to a phenomenon known as catastrophic forgetting---where the model ``forgets'' much of the original information from its original training data \citep{Luo2023forgetting}.
}

Designing theory-grounded agents helps us improve fidelity, but their effectiveness ultimately depends on what the underlying LLM already knows about the world.
In practice, this means that some domains will be more conducive to reliable simulation than others.
Certain types of settings may therefore inspire greater confidence that AI simulations will be effective---specifically, those we would expect to be well-represented in the LLM's training data.
This follows a simple machine learning principle: models are most accurate in domains richly represented in their training data.
Indeed, evidence shows that LLMs perform far better on problems and questions drawn from well-covered settings \citep{EmbersMcCoy2024}.

Perhaps unsurprisingly, two of the largest experimental papers in terms of participant sample size and number of distinct experiments demonstrating that LLMs have broadly applicable predictive power in novel settings focus on domains that are likely to be richly represented in digital text: laboratory experiments in psychology \citep{binz2025centaur} and digital marketing experiments \citep{hewitt2024predictingLLM}. 
Although we cannot know the boundaries of what is in the training corpus without open models---and even then, the sheer volume of data may obscure clear boundaries---trusting results from simulations of settings well-captured by text on the internet is likely a sound heuristic.

\section{Conceptual issues} 
\label{sec:conceptual_issues}
Critiques of \emph{Homo silicus} question the suitability of using LLMs for social science.
One potential response mirrors the argument of \cite{friedman1953methodology} that the realism of \emph{Homo economicus} and our assumptions more generally do not matter, and instead we should evaluate an approach on whether it can generate useful results.
The proof of the pudding is in the eating, and ``eating'' for our purposes is whether AI simulations can help us to expand the frontiers of human knowledge more quickly.
For example, suppose that the primary use of \emph{Homo silicus} is to simulate experiments before we try them in the real world, and this method is adopted. 
In that case, these debates are somewhat moot.
Nevertheless, it is helpful to consider non-Friedman responses to common critiques of AI simulations.

\subsection{The ``garbage in, garbage out'' critique}
\label{ssec:garbage_in_garbage_out}
One critique of using LLMs for social science is that their training data is too large to be carefully curated, and this may create a ``garbage in, garbage out'' problem.
This critique rests partly on the incorrect notion that the responses of LLMs are a sort of weighted average.
However, LLMs are more like random number generators than estimators. 
If an LLM is trained on millions of people reporting random draws from $U[0,1]$, it would not always respond with approximately $0.5$, but rather it would be more or less equally likely to return any number in $[0,1]$.
LLM developers also have strong incentives to curate the training data carefully~\citep{brown2020language, llama3Meta2024}, and post-training methods act as an additional layer of filtering.

Even great data may disproportionately represent the highly selected pool of ``humans creating public writing.''
One advantage economists have in using LLMs is that they tend to pose questions that place few demands on the sample.
We do not think of demand curves sloping downward as a ``Western, Educated, Industrialized, Rich, and Democratic'' phenomenon but rather as a result of rational goal-seeking that all humans engage in.\footnote{Although preliminary work shows how LLMs may effectively model non-WEIRD populations~\citep{capra2024llms}.}
The willingness of economists to use undergraduates at elite four-year universities for laboratory experiments is partially a convenience---but also consistent with a disciplinary point of view we share with psychologists that it is not likely to matter much.

A more subtle problem is that LLMs are trained on what humans choose to say---not on what they choose to do.
Economists have historically taken a dim view of the economic content of mere statements rather than behaviors. 
This would seem to be a damning critique of a model literally trained on statements.
However, this ``stated versus revealed preference'' critique is only superficially persuasive.
As we have discussed and illustrated in Figure~\ref{fig:diagram_knowledge}, the internet contains both implicit and explicit sources of social information.
The training data is not millions of lines of people lying about their reservation values in bargaining scenarios. 
Instead, much of it is text about people reasoning how to approach various economic questions, including ``stage whispers'' about their true intentions, explaining how they would deal with a situation, or recording economic decisions such as purchases and ratings.\footnote{The parent of one of the authors runs a construction company and has to negotiate constantly. 
He claims that one of his useful negotiating skills is the ability to read upside down because many people will write their reservation value on a piece of paper they have in front of them (often underlined). 
By putting this text in a paper on the public internet, this tiny piece of human reasoning about economic life is now available for LLMs to learn.} 

\subsection{``Memorization'' and ``performativity'' problems}
\label{ssec:memorization_performativity}
With billions of parameters and a massive training corpus, one worry is that LLMs are parroting back to us what they have ``read'' somewhere in their training data.
This view is not entirely correct. 
First, it is inconsistent with the fact that LLMs can ``hallucinate'' and make up new facts.
Second, our findings in Section~\ref{sec:experiments} contradict this view: the AI and human subjects' responses differ in scenarios that are in the training data.
Moreover, memorization is not intrinsically bad.

Memorization would be problematic if the LLM's predictions were so ``shallow'' as to be brittle.
To make an analogy, consider an economics student studying for a game theory exam.
A bad student has memorized that \{Defect, Defect\} is the unique pure strategy Nash equilibrium for the prisoner's dilemma game, and then selects that prediction when it appears on the midterm.
This shallow memorization is problematic because it does not generalize to games with different payoffs or action spaces.
A better student would memorize that a Nash equilibrium is a strategy profile where every player best-responds to every other player, and would be able to apply this concept to identify equilibria in novel settings.
The usefulness and flexibility of LLMs suggests that their predictions are often more like the good student's.

Another related concern is a potential ``performativity'' problem, in the sense that AI agents may behave following the theories and empirical results in their training data~\citep{performativity2007}.
However, as with the ``good'' game theory student, performativity may be a desirable feature of AI simulations---assuming the underlying theory is correct.
More generally, a critical view of memorization and performativity overshadows their important benefits. 
If our goal is to have a ``realistic enough'' model of human behavior, rather than to make unequivocal inference about human behavior, and we believe that the cumulative body of social science is useful in predicting actual human behavior, then internalizing and applying these facts is beneficial. 

\subsection{Black boxes with incoherent world models}
We do not fully ``understand'' the behavior of LLMs today, including how training data, model architectures, and prompts affect model outputs.
A deeper understanding would certainly be useful, but it is not necessary to use LLMs for social science.
To do economics, we do not have to study neurons and parts of the brain.
Instead, we advance our body of knowledge by studying how humans respond in different scenarios and by developing theories, one study at a time. 
Similarly, what matters is not that we fully understand how LLMs work, but rather that they are systems created for some understandable purpose with an explicit goal of optimization.
As \cite{simon1996} notes, ``sciences of the artificial'' can usefully abstract away from the micro-details of construction so long as the created object is viewed as trying to maximize something subject to the constraints of the environment.

A related concern is that LLMs may have incoherent world models, which can undermine their credibility.
An interesting case in point is \citet{vafa2024world}.
They train a language model to predict paths between points in New York City based on rides from taxi drivers.
To do this, they label every single intersection in the city with a unique identifier.
Each ride is then tokenized as an initial intersection, an ending intersection, and a series of actions---``left,'' ``right,'' ``straight,'' etc.---between them.
When trained on thousands of these rides, the model can predict the correct sequence of actions to get between hold-out sets of starting and ending intersections with high accuracy.
Yet, the authors show that the inferred map of New York City omits streets, invents others, and the predictions are far less accurate when street detours are introduced.
The language model's world model is brittle.
It is not explicitly designed to be a map of the city.\footnote{\cite{vafa2025planets} offer another related context focusing on Newtonian mechanics.
Some points largely ignored by both papers are the now-well-established scaling-and-breadth effects: LLMs trained on broad, heterogeneous tasks tend to have the best performance across the board \citep{kaplan2020scalinglaws,brown2020language,sanh2022multitask,2024gpt4technicalreport}. 
It is unclear how the most capable LLMs suffer from the incoherencies in \cite{vafa2024world}.}

While a perfectly coherent world model would be a boon to AI simulations, it is not strictly necessary to have credible predictions in many contexts.
Effective simulations can rely on what the LLM was designed to do---follow instructions.
Consider an LLM being used to predict how humans might navigate some city.
Given a perfect GPS, this would be trivial.
However, the LLM might still navigate well if given a broadly applicable set of instructions and updated information about the immediate environment.
If it were endowed with \emph{``Stop at stop signs. Stay on the right side of the road. Follow the speed limit.''} and intermittent updates about the immediate environment, it could move around reasonably. 
Navigating with those instructions is a far simpler task---and requires a far simpler world model---than reconstructing the entire street grid from scratch, and it is robust to many small perturbations.\footnote{This perspective may also shed light on why many AI simulations in the literature have such poor fidelity.
Much of this research focuses on the prompts as simple social and demographic traits \citep{1000agents2024park,opinions2023Santukar,WhichHumans:2023,rottger2024political}.
Such ``instructions'' like \emph{``respond as a 30-year-old male''} or \emph{``respond like a professor from MIT''} assume the LLM can coherently represent how those traits interact with the strategic setting.
When its world model is shaky, as may often be the case for complex constructs like the interactions of human identities, the simulation may deteriorate or default to caricatures of the target persona \citep{Cheng2023caricature}.}

\subsection{Are these ``just simulations?''} 
\label{ssec:just_simulations}
An objection to AI-based experiments is that they are just simulations or agent-based models.
Economists generally take a dim view of simulation-based approaches because the researcher is both judge and jury: you program the agents and then see what they do.
Rather than ``What would \emph{Homo economicus} do?'' simulation-based approaches often ask ``What would this model [that does what I tell it to do] do?''---a far less scientifically interesting question. 
This is arguably why \cite{schelling1971dynamic} is the exception that proves the rule.
His model was so simple and obvious that readers knew there was no card up his sleeve, no trick to ensure the surprising emergent phenomenon.

In contrast to conventional agent-based models, \emph{Homo silicus} is not under direct researcher control. 
We can influence \emph{Homo silicus} with endowments of beliefs, experiences, and so on (see Section~\ref{sec:experiments}).
But we remain constrained by the fact that what determines behavior is the underlying model, not our direct programming.

\emph{Homo silicus} also gives researchers a much greater degree of flexibility than agent-based models in the past.
With LLMs, we can map preferences, beliefs, and so on, to purposeful actions in the simulated social world; 
we can ``program'' complex simulations only using natural language; 
and we can often parse the LLMs' responses to study the reasoning behind their choices.
These factors expand greatly the scope of AI simulations.

The flexibility of \emph{Homo silicus} makes it less susceptible to one of the core criticisms of both conventional agent-based models and economic theory itself: the Lucas critique \citep{Lucas1976Critique}.\footnote{This may help explain why agent-based models have had relatively little uptake in mainstream economics.}
At its core, the Lucas critique emphasizes that behavioral rules guiding agents cannot be treated as fixed parameters but as endogenous responses shaped by the prevailing policy environment. 
LLM-based agents mitigate this concern because they can engage in flexible reasoning about environmental changes rather than following hard-coded behavioral rules. 

\subsection{It is easy to ``prompt hack'' AI experiments}
The technical requirements of AI simulations are minimal.
This low barrier to entry, however, comes with the risk of researchers iterating through many prompts until they get a desired response and report only favorable results.
This is conceptually similar to the problem of p-hacking in traditional experimental work.

There are two easily implementable solutions to this problem.
First is to require researchers to present their results under many different partially controlled permutations of the prompt as a robustness check.
For example, in the \cite{kahneman1986fairness} experiment, we repeated the experiment with different temperatures, translated the prompt into other languages and then back to English, and varied the context of the experiment.
These manipulations varied the data-generating process in a way that is partially but not entirely within our control.
The second solution, which we have implemented in all experiments throughout this paper, is to make all simulation code public.
Then anyone can inspect the exact data-generation process underlying any analysis.
Indeed, this transparency and portability are major advantages of AI relative to human subjects.
Replicating experiments often only requires a few dollars and the click of a button.

Both solutions require little of the researcher beyond a willingness to share their work---a willingness they should have anyway.
Coding up alternative simulations is straightforward.
It requires a few minutes of additional prompting and some minor code changes.
Sharing results is also easy.
Frontier LLMs can clean up code syntax and provide comprehensive documentation for an entire codebase with a single prompt.

\subsection{Causal inference with AI simulations is challenging}
\label{sec:causal}
Two features of LLM experiments ostensibly make causal analysis difficult.
First, an LLM is a single model, and so it would seem that we can only get one ``observation.''
Unlike the one \emph{Homo economicus} that is rational, there are many \emph{Homo silici}.
We demonstrated in Section~\ref{sec:experiments} that we can induce LLMs to play different agents via prompts, leading to varying responses.\footnote{\cite{argyle2022out} make this point in their perfectly titled ``Out of one, many'' paper---there is not a single LLM but rather a model capable of being conditioned to take on different personas that respond realistically.}
This agent ``programming'' is not unlike the experimental economics practice of giving a subject a card that says their marginal cost of producing a widget is 15 tokens.
Additional sources of variation may come from setting the temperature parameter and from using different LLMs.

Which set of \emph{Homo silici} to use depends on the researcher's question of interest.
Researchers should not just use off-the-shelf LLMs to conduct simulations, but rather construct a set of agents relevant to their domain of study.
This may require matching simulations to existing data or grounding the agents on existing theories.

A second challenge concerns identification.
Even with perfect randomization, all else may not be equal when comparing treatment groups in AI experiments \citep{Gui2023Causal}.
The core challenge is that editing a prompt to change one factor may inadvertently cause other factors to change; the experiment may then not manage to isolate a single cause.
However, such a problem is a question of external validity---whether a given randomized manipulation generalizes to environments with other downstream variables, variables that may influence outcomes in ways the original controlled experiment did not capture.

External validity is a problem for experiments both with human and \aisub{s}.
Consider the prompt parameterized by the wage asks of the two applicants in our dishwashing experiment. 
The experimental design iterated over combinations of the two wage asks, with an externally imposed minimum wage forcing up the wage asks when they were below \$15/hour.
We first ran the experiment without a reference wage.
When we repeated the experiment with a reference wage of \$12, the effect of the minimum wage on both the wage and experience of the hired worker increased.

The critique is an explanation for this difference.
Without a reference wage, the LLM may impute values for the reference wage as a function of the wage asks learned during training.
And if the LLM's response distribution changes with the reference wage, then treatment assignment could influence the potential outcomes of the hired worker's wage and experience.
Instead of the experiment generating the LLM's response based on the hiring prompt with no reference wage, the experiment is actually based on the hiring prompt with an imputed reference wage that depends on the wage asks.
Conversely, when the reference wage is \$12, the reference wage is specified, so there is no value to impute.

Now, suppose we repeated the \cite{laborHorton2025} experiment with people---with and without a reference wage.
It seems plausible, even likely, that if unspecified, the hiring manager would presume some sort of reference wage for applicants based on their wage asks.
As with the LLM experiment, human subjects may impute values for unspecified information that depend on the treatment and their past experiences.
This imputation may affect the potential outcomes.

Stepping back, both the AI and the hypothetical human subjects' experiments are perfectly randomized.
We can always estimate the downstream causal effects of such exogenous manipulations.
But if we hope to generalize these results to other settings, to claim that the results are externally valid, we must be careful; unspecified information may be relevant to the causal relationships under investigation.

In human-subjects experiments, we deal with questions of external validity in a few ways.
If the experiment is executed in the field (e.g., in a labor market with real employers), it operates in the exact environment of interest, and all possible information is, by definition, specified.
This is not the norm.
Many experiments are much less naturalistic---they are conducted in a lab or a lab-in-the-field and surely only specify some of the information that might be relevant to the outcome of interest \citep{Levitt2007What}.
In these cases, it is incumbent upon the researcher to justify the generalizability of the results \citep{Camerer2015LabField}.
They can do this through theory, robustness checks, additional experiments, etc.
We can use AI simulations to help here, to think deeply about the potential sparsity in our experiments and how it might affect the results---just as we do with human-subjects experiments.
We can run robustness checks and, most importantly, run additional simulations quickly at negligible marginal cost \citep{shahidi2025coasean}.
\emph{Homo silicus} can provide a powerful check on the assumptions researchers may often take for granted and help us identify possible threats to generalizability.

\section{Conclusion} \label{sec:conclusion}

This paper reports the results of several experiments using dozens of LLMs as simulated experimental subjects.
We recover core qualitative patterns observed in the analogous human-subjects experiments and surface deviations that are themselves likely informative for future research.
These demonstrations highlight our essential conceptual contribution: \emph{Homo silicus} is best understood as theory in flexibly executable form.
AI simulations can also be a fast, low-cost way to explore ideas, stress-test experimental designs, and calibrate power calculations before committing to costly human data collection.

We make a practical contribution by releasing open-source software that standardizes the design and implementation of AI experiments.
It allows researchers to switch in any model to run an experiment, making it easy to update results even as models evolve.
All experiments in this paper are now button-push reproducible, with extensive documentation.
Our view is that open, reproducible research can be facilitated by pushing the costs of replication, transparency, and distribution toward zero.

In terms of future work, the most pressing task is to better understand when AI simulations are high-fidelity: to sketch the families of settings and assumptions where they track \emph{Homo sapiens}, and to be candid about where they do not.
This missing map is the key limitation today.
However, there is already progress on this front.
For now, we can say that AI simulations are most informative when mechanisms can be written as clear instructions, the domain is well represented in digital text, and researchers can benchmark on related tasks with human data when possible.
As we improve our mapping, future opportunities will emerge.
One particularly promising direction is automated scientific loops: systems that propose hypotheses, instantiate agents, run and score simulations, and iterate \citep{Manning2024Automated}.
Such systems could dramatically accelerate research productivity.

\singlespacing
\bibliographystyle{aer}
\bibliography{homo_silicus.bib}

\newpage \clearpage 

\doublespacing
\appendix

\renewcommand{\thefigure}{A\arabic{figure}} 
\setcounter{figure}{0}  

\renewcommand{\thetable}{A\arabic{table}} 
\setcounter{table}{0}  

\section{Figures from the original draft} \label{sec:original_figures}

\begin{figure}[h]\caption{\cite{charness2002understanding} Simple Tests choices by model type and endowed ``personality''} \label{fig:original_charness_rabin}
  \centering
  \begin{minipage}{0.99\textwidth}
    \includegraphics[width = \linewidth]{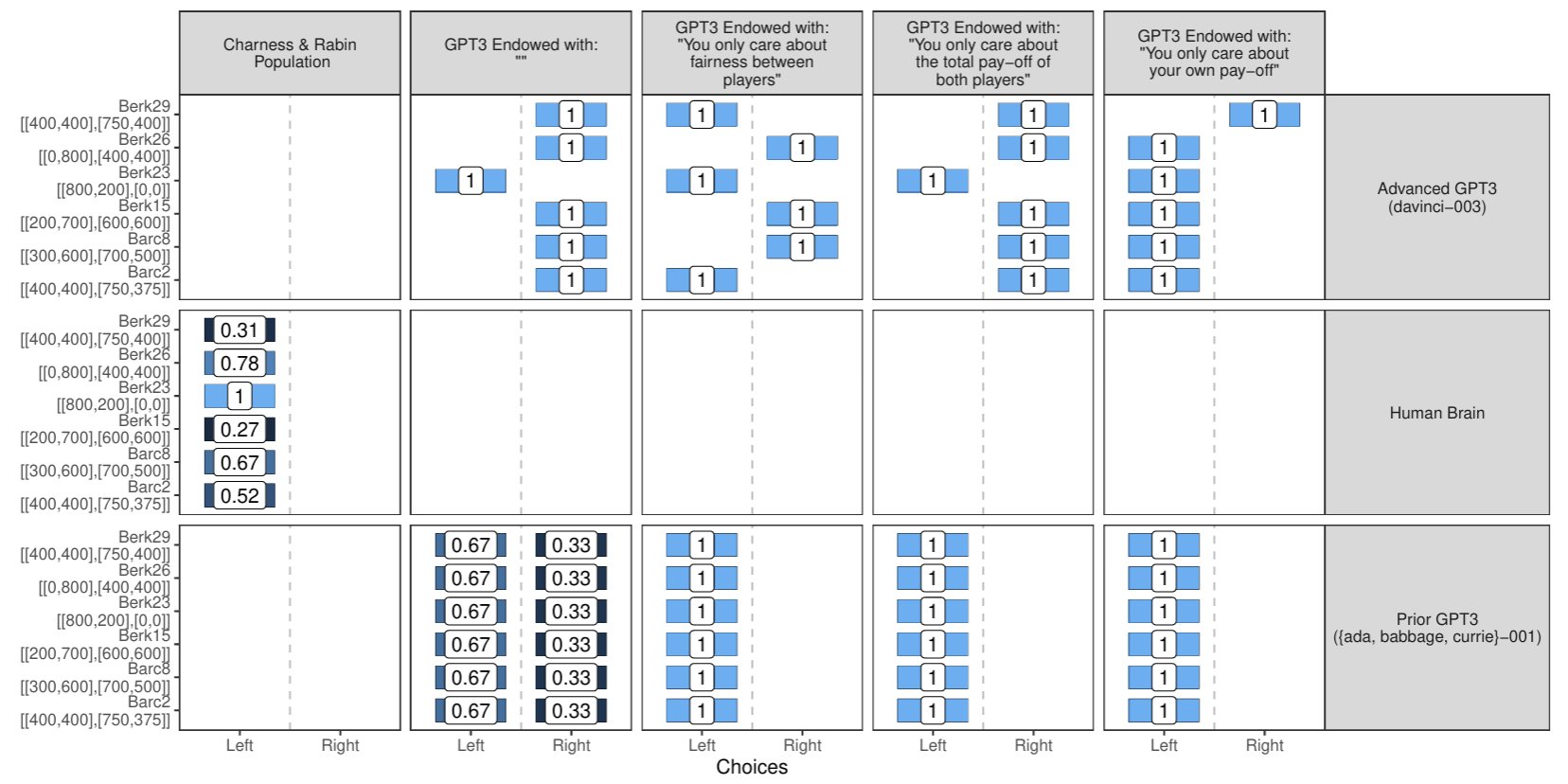}
{\footnotesize \\
  \emph{Notes:} This shows the fraction of \aisub{s} choosing each option, by framing.
}
\end{minipage} 
\end{figure}

\begin{figure}[h]
  \caption{\cite{kahneman1986fairness} price gouging snow shovel question, with endowed political views} \label{fig:original_kkt}
  \centering
  \begin{minipage}{0.99\textwidth}
    \includegraphics[width = \linewidth]{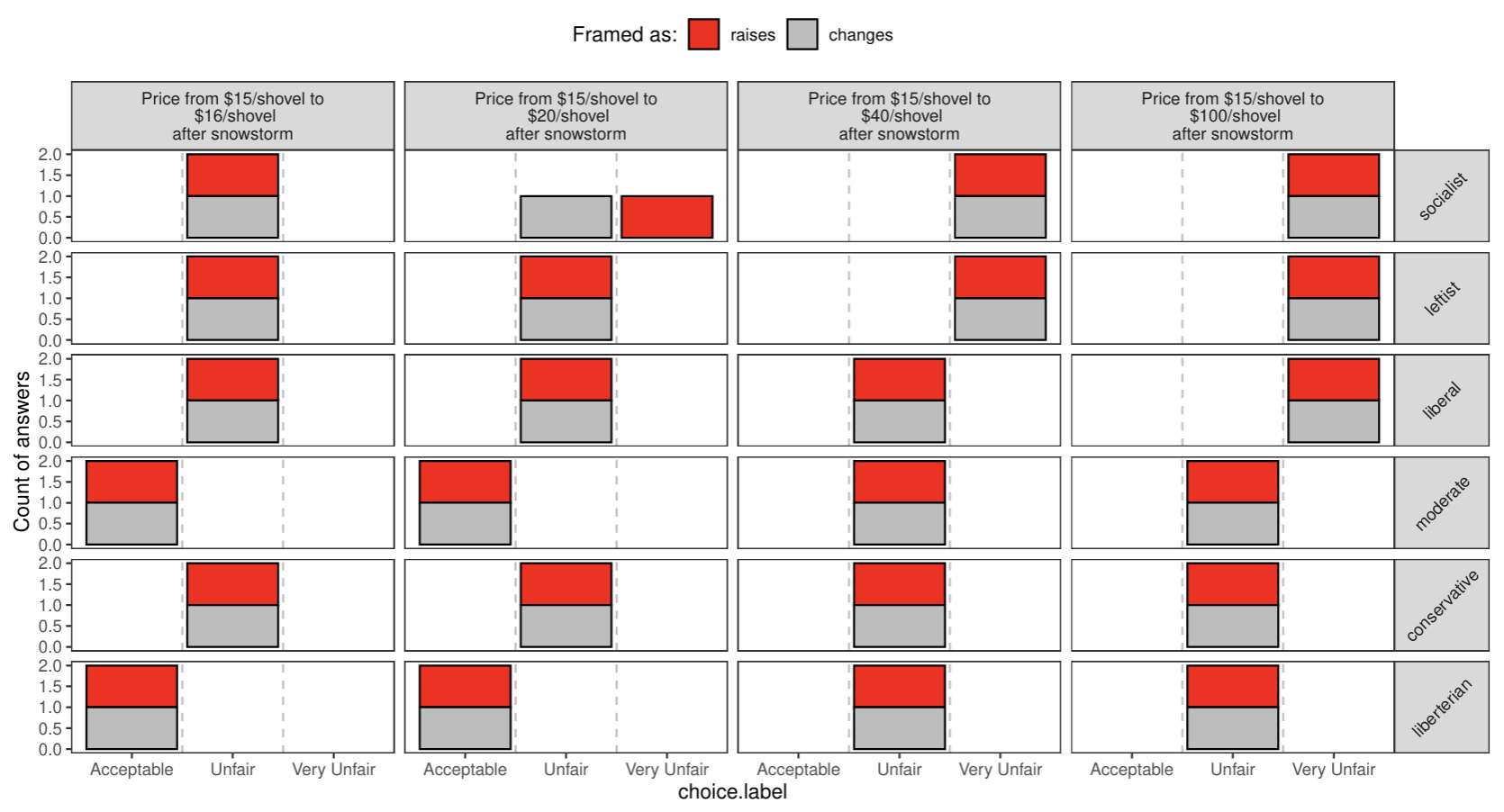}
{\footnotesize \\
  \emph{Notes:} This shows the fraction of \aisub{s} choosing each opinion, by scenario.
}
\end{minipage} 
\end{figure}

\begin{figure}[h]
  \caption{Distribution of preferred car safety budgets, by status quo framing} \label{fig:original_zeckhauser}
  \centering
  \begin{minipage}{0.99\textwidth}
    \includegraphics[width = \linewidth]{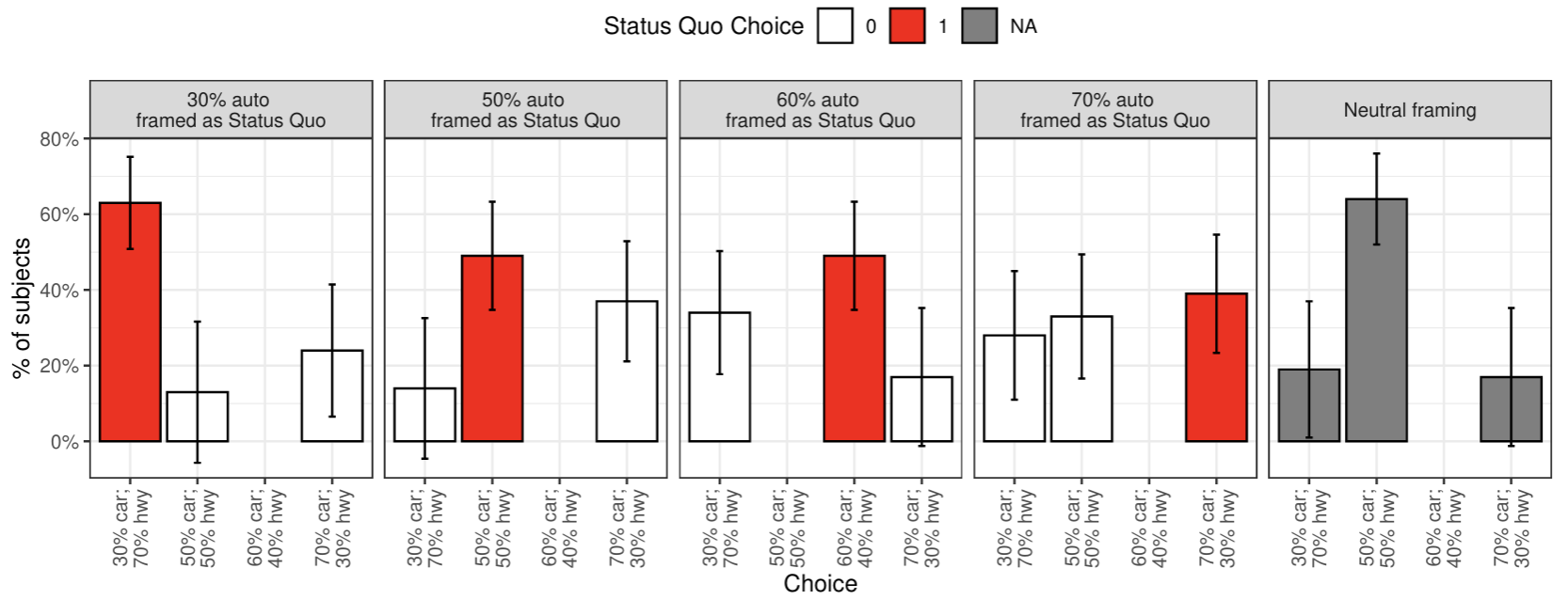}
{\footnotesize \\
  \emph{Notes:} This shows the fraction of \aisub{s} choosing each option by framing.
}
\end{minipage} 
\end{figure}

\begin{table}[h]
  \caption{Effects of minimum wage on observed wage and hired worker attributes} \label{tab:original_horton}
  \centering
  \begin{minipage}{0.99\textwidth}
    \includegraphics[width = \linewidth]{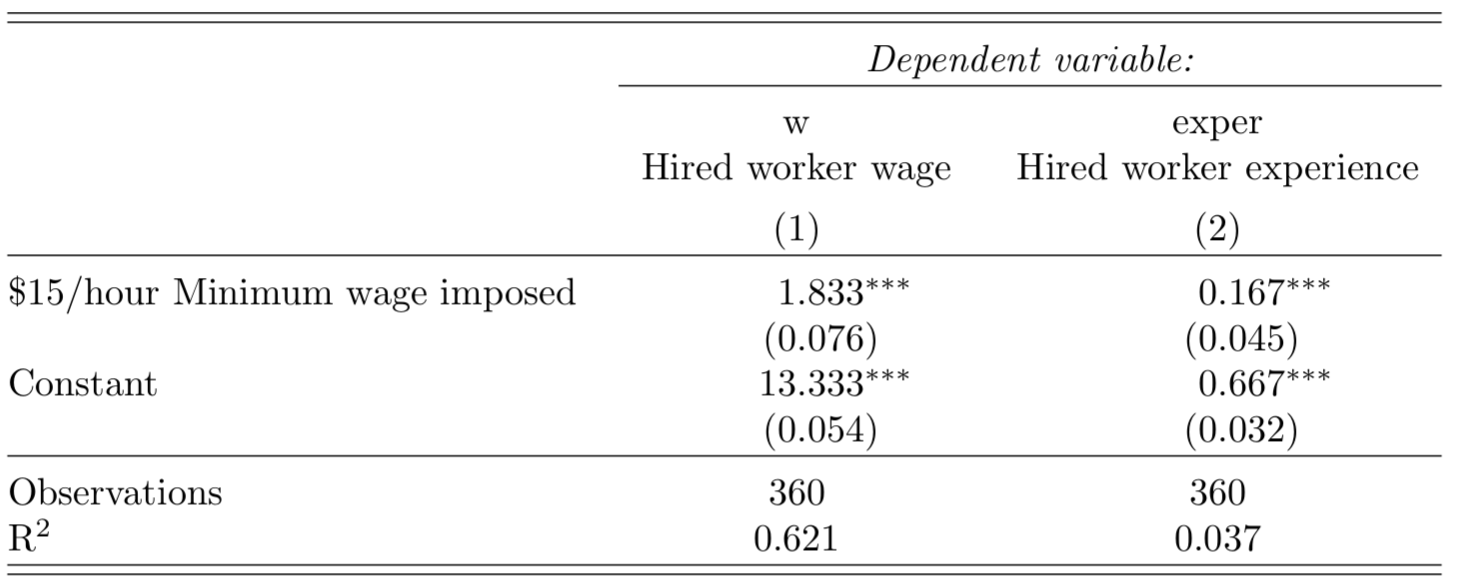}
{\footnotesize \\
  \emph{Notes:} This reports the results of imposing a minimum wage on the (1) hired worker wage and (2) hired worker experience. 
}
\end{minipage} 
\end{table}

\newpage \clearpage

\section{A brief primer on large language models}
\label{app:primer}
LLMs are neural networks with complex architectures and large numbers of parameters, estimated (trained) using massive amounts of text.
Broadly speaking, they function as follows.
Given an input prompt, an LLM generates a response by first predicting the most likely next word, and then using the prompt and the words predicted so far to predict subsequent words.\footnote{Technically, LLMs use ``tokens,'' which can be sentences, words, parts of words, or characters.}

Training an LLM consists of a ``pre-training'' phase and a ``reinforcement learning with human feedback'' (RLHF) phase. 
In the pre-training phase, the LLM learns to predict the next word in sequences of text in its training data.
A pre-trained LLM estimates a conditional probability distribution over its training data; the exact estimated distribution depends on design choices such as the architecture and hyperparameters of the LLM.

After pre-training, the model may undergo one or more post-training stages.
The most common post-training step is reinforcement learning with human feedback (RLHF), in which human evaluators rate model outputs, their judgments are used to train a reward model, and the LLM's parameters are adjusted so that its responses receive higher predicted rewards.
Other post-training steps include supervised fine-tuning (or instruction-tuning), where human-written examples ``teach'' the model to follow prompts or exhibit a desired style.
In the RLHF phase, the pre-trained LLM is trained further using human feedback.
First, human evaluators prompt the LLM and evaluate its responses.\footnote{For example, the human evaluators may rate the responses on a scale of 1 to 10, depending on how well a response answers a question or matches a particular style of writing.}
The human evaluation data is then used to estimate a separate ``reward model,'' which predicts human feedback given a prompt and an LLM response.
The LLM's parameters are updated through a form of reinforcement learning so that its responses receive favorable feedback from the reward model.
The fully trained LLM's responses are hence optimized toward both accurately predicting the likely next words in sequences of text and receiving high feedback scores from the reward model.
An LLM can always undergo further fine-tuning to make its responses more consistent with a particular persona or writing style.
  
Users can adjust how the LLM responds to input prompts by changing its ``temperature'' parameter.
The temperature parameter controls how the LLM samples from its response distribution.
At temperature zero, the LLM is deterministic: it always outputs the highest-probability response.
Increasing the temperature makes the output stochastic and the response distribution more uniform.

\section{Additional Figures and Tables}\label{sec:appendix_figures}

\begin{figure}[h]
\begin{center}
\caption{Example screenshot of linked Jupyter notebooks for running AI simulations}
\includegraphics[width=0.95\textwidth]{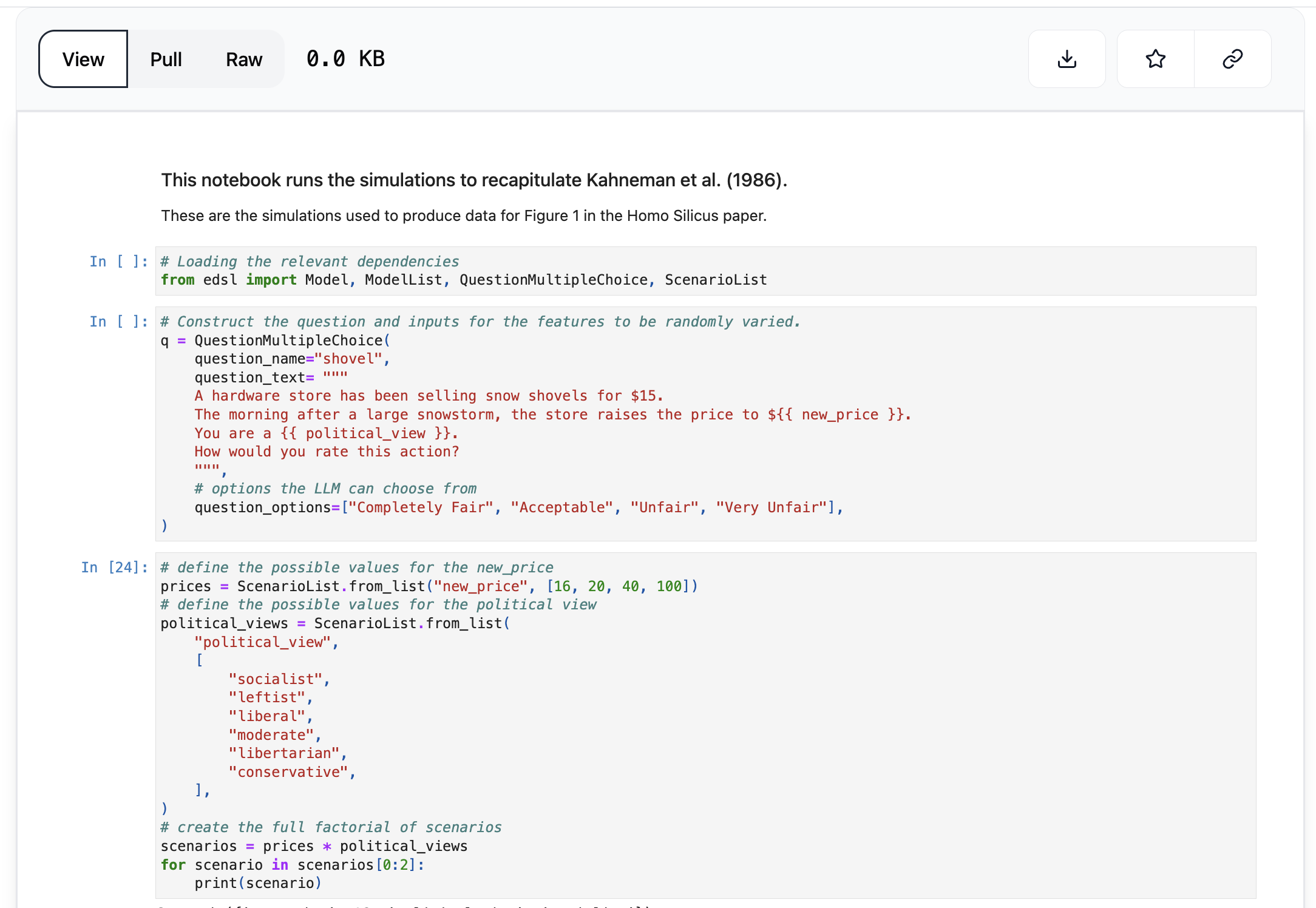}
\label{fig:edsl_screenshot}
\end{center}
\begin{footnotesize}
\begin{singlespace}
\emph{Notes:} This figure shows a screenshot of the Jupyter notebook for the \cite{kahneman1986fairness} experiment.
All notebooks in this paper share a similar structure, with instructions and comments throughout the code.
Each also includes links to share or download the notebooks so others can easily replicate the experiments or explore their own ideas.
\end{singlespace}
\end{footnotesize}
\end{figure}  

\begin{table}[h!] 
\begin{center}
\begin{minipage}{1 \linewidth}                          
\caption{Estimates of the effects of political beliefs and price hike levels on \aisub{s}' fairness assessments in four permutations of the \cite{kahneman1986fairness} experiment.}  
\label{tab:kkt}     
\centering                                                        
\small
\begin{tabular}{@{\extracolsep{.5pt}}lcccccc} 
\\[-1.8ex]\hline 
\hline \\[-1.8ex] 
 & \multicolumn{6}{c}{\textit{Dependent variable:}} \\ 
\cline{2-7} 
\\[-1.8ex] & \multicolumn{6}{c}{Choice as Numeric from 1 (Very Unfair) to 4 (Completely Fair))} \\ 
\\[-1.8ex] & (1) & (2) & (3) & (4) & (5) & (6)\\ 
\hline \\[-1.8ex] 
 $\Delta$ Price & $-$0.006$^{*}$ &  & $-$0.006$^{*}$ & $-$0.004$^{*}$ & $-$0.009$^{*}$ & $-$0.007$^{*}$ \\ 
  & (0.001) &  & (0.0002) & (0.0003) & (0.0002) & (0.0003) \\ 
  & & & & & & \\ 
 Socialist &  & $-$1.438$^{*}$ & $-$1.437$^{*}$ & $-$0.598$^{*}$ & $-$1.005$^{*}$ & $-$1.172$^{*}$ \\ 
  &  & (0.030) & (0.027) & (0.038) & (0.027) & (0.030) \\ 
  & & & & & & \\ 
 Leftist &  & $-$1.383$^{*}$ & $-$1.383$^{*}$ & $-$0.536$^{*}$ & $-$1.065$^{*}$ & $-$1.113$^{*}$ \\ 
  &  & (0.030) & (0.027) & (0.038) & (0.027) & (0.030) \\ 
  & & & & & & \\ 
 Liberal &  & $-$0.733$^{*}$ & $-$0.733$^{*}$ & 0.115$^{*}$ & $-$0.453$^{*}$ & $-$0.653$^{*}$ \\ 
  &  & (0.030) & (0.027) & (0.038) & (0.027) & (0.030) \\ 
  & & & & & & \\ 
 Conservative &  & 0.307$^{*}$ & 0.307$^{*}$ & 0.008 & 0.422$^{*}$ & 0.522$^{*}$ \\ 
  &  & (0.030) & (0.027) & (0.038) & (0.027) & (0.030) \\ 
  & & & & & & \\ 
 Libertarian &  & 0.860$^{*}$ & 0.860$^{*}$ & 0.847$^{*}$ & 1.223$^{*}$ & 1.205$^{*}$ \\ 
  &  & (0.030) & (0.027) & (0.038) & (0.027) & (0.030) \\ 
  & & & & & & \\ 
 Constant & 2.462$^{*}$ & 2.687$^{*}$ & 2.860$^{*}$ & 2.352$^{*}$ & 2.734$^{*}$ & 2.576$^{*}$ \\ 
  & (0.025) & (0.021) & (0.020) & (0.029) & (0.020) & (0.022) \\ 
  & & & & & & \\ 
\hline \\[-1.8ex] 
Experiment & Baseline & Baseline & Baseline & Translated & Variations & Adversarial \\ 
Observations & 2,400 & 2,400 & 2,400 & 2,397 & 2,400 & 2,400 \\ 
R$^{2}$ & 0.043 & 0.802 & 0.846 & 0.460 & 0.834 & 0.823 \\ 
\hline 
\hline \\[-1.8ex] 
\end{tabular}

\end{minipage}
\end{center}
\begin{footnotesize}
\begin{singlespace}
\vspace*{-0.15in}
\emph{Notes:}
This table reports regressions where the independent variables are the snow shovel dollar price change relative to its original \$15 price, and the \aisub{s}' political beliefs.
We use ``Moderate'' as the reference class.
The dependent variable is the fairness assessment of the subjects, which we measure as a continuous variable ranging from 1 (``Very Unfair'') to 4 (``Completely Fair'').
The $\Delta$ Price coefficient represents the change in fairness rating (on the 1--4 scale) for each \$1 increase in price above the \$15 baseline.
Columns (1-3) show the results for the baseline experiment;
Column (4) for the translated prompts experiment;
Column (5) for the prompt variations experiment; and
Column (6) for the adversarial prompt experiment.
\starlanguage{}
\end{singlespace}
\end{footnotesize}
\end{table}

\end{document}